\documentclass{article}
\usepackage[pdftex]{graphicx}

\begin{document}

\begin{center}
SUPER-RADIANT DYNAMICS, DOORWAYS, AND RESONANCES \\
IN NUCLEI AND OTHER OPEN MESOSCOPIC SYSTEMS \\
\vspace{0.5cm}
Naftali Auerbach$^{1,2}$ and Vladimir Zelevinsky$^{2,3}$\\
\vspace{0.5cm}
$^{1}${\sl School of Physics and Astronomy, Tel Aviv University, Tel Aviv, 69978, Israel \\
$^{2}$Department of Physics and Astronomy, Michigan State University, \\
East Lansing, Michigan 48824-2320, USA\\
$^{3}$National Superconducting Cyclotron Laboratory, Michigan State University,
East Lansing, Michigan 48824-1321, USA}
\end{center}

\begin{center}
{\small {\bf Abstract}}
\end{center}

{\small The phenomenon of super-radiance (Dicke effect, coherent spontaneous
radiation by a gas of atoms coupled through the common radiation field) 
is well known in quantum optics. The review discusses similar physics that
emerges in open and marginally stable quantum many-body systems. In the presence
of open decay channels, the intrinsic states are coupled through the continuum.
At sufficiently strong continuum coupling, the spectrum of resonances undergoes
the restructuring with segregation of very broad super-radiant states and trapping
of remaining long-lived compound states. The appropriate formalism describing this
phenomenon is based on the Feshbach projection method and effective non-Hermitian
Hamiltonian. A broader generalization is related to the idea of doorway states
connecting quantum states of different structure. The method is explained in detail
and the examples of applications are given to nuclear, atomic and particle physics.
The interrelation of the collective dynamics through continuum and possible
intrinsic many-body chaos is studied, including universal mesoscopic conductance
fluctuations. The theory serves as a natural framework for general description of 
a quantum signal transmission through an open mesoscopic system.}  

%\end{document}

\newpage
%\end{document}

\begin{enumerate}

\item Introduction

\item Origin of super-radiance

\begin{itemize}
\item Original idea
\item Transport through a lattice
\end{itemize}

\item Open quantum system

\begin{itemize}
\item Projection method and effective non-Hermitian Hamiltonian
\item Effective Hamiltonian and reaction theory
\item Emergence of resonances
\item Overlapping resonances and super-radiance
\end{itemize}

\item Nuclear continuum shell model

\begin{itemize}
\item Two-level example
\item Realistic calculations
\item Example: Oxygen isotopes
\item Some technicalities

%\begin{itemize}
%\item Oxygen isotopes
%\end{itemize}

\end{itemize}

\item Doorway states

\begin{itemize}
\item Dynamics based on doorway states

\begin{itemize}
\item The case of a single doorway
\item When is this approach valid?
\item Isobaric analog resonances
\end{itemize}
\item Giant resonances
\item Several channels and intermediate structure
\item Highly excited doorway state
\item Additional comments
\end{itemize}

\item Other applications

\begin{itemize}
\item Atomic and molecular  physics

%\begin{itemize}
%\item{Autoionizing states}
%\end{itemize}

\item Intermediate and high energy physics

\begin{itemize}
\item The $\bar{N}N$ states
\item The $\Delta$-nucleon system
\item High energy phenomena
\end{itemize}

\end{itemize}

\item Quantum chaos in an open system

\begin{itemize}
\item Many-body quantum chaos
\item Statistics of resonances
\item Statistics of resonance widths
%\begin{itemize}
%\item Isolated resonances (weak continuum coupling)
%\item Overlapped resonances (strong continuum coupling)
%\end{itemize}
\item Role of intrinsic chaos
\end{itemize}

\item Signal transmission through a mesoscopic system

\begin{itemize}
\item Common physics of mesoscopic systems
\item Statistics of cross sections
\item Fluctuations of cross sections
\item From Ericson to conductance fluctuations
\end{itemize}

\item Conclusion and outlook

\end{enumerate}

%\end{document}

\newpage

\section{Introduction}

More than half a century ago Robert Dicke published an article where he
predicted the phenomenon of {\sl super-radiance} \cite{dicke54}. This phenomenon
was indeed experimentally observed in an optically pumped gas \cite{skribanowitz73}
and later studied in detail and widely used in quantum optics
\cite{andreev93,benedict96} as well in physics of plasmas and beam physics
\cite{menshikov99}. The super-radiance was also observed with
the free-electron laser \cite{watanabe07} and with atomic matter waves,
including Bose-Einstein condensate \cite{kaiser01}. The quantum-mechanical
essence of super-radiance in the original formulation is relatively simple.
Two-level systems (``atoms" or ``qubits") can be coupled even in the absence
of their direct interaction. The coupling proceeds through the common
radiation field. The migrating resonant excitation combines the atoms
in a collective capable of releasing the burst of radiation as a whole.
The phenomenon therefore is called {\sl coherent spontaneous radiation}
that sounds as an oxymoron since traditionally the spontaneous radiation
was considered as being incoherently produced by individual atoms.

In the end of the nineteen eighties it was understood that the physics
underlying super-radiance is much more general and can find broad
applications in various regions of the quantum world. The necessary ingredient
is the set of unstable quantum states with the same exact quantum numbers.
The states should be capable of decaying into a common decay channel
(or few channels). This creates an analog of their coupling through the common
``field of radiation", or common continuum; the coupling becomes effective
when the widths of individual adjacent states grow and overlap. The system of
overlapping resonances undergoes a sharp restructuring forming super-radiant
states, one for each common open channel. In the ideal version of quantum
optics, when the wavelength of radiation significantly exceeds the size
of the system, the identical atoms have identical levels and widths which
amplifies the effect. In a general case, the levels are not degenerate,
and the super-radiant transition can happen only when the typical widths
are of the order of typical spacings between the unstable levels.

The mechanism of super-radiance is omnipresent being simply a consequence of
quantum mechanics. It was seen in the problems of quantum chemistry, atomic and
molecular physics. In a pure form it was discovered in nuclear physics in
consideration of the decay of giant resonances \cite{kleinwachter85} when the
numerical simulation clearly indicated the appearance of very broad resonances
for each open decay channel when the continuum coupling exceeds a critical
strength. At the same time the remaining resonances are getting more narrow
signaling the existence of {\sl trapped states} with long lifetime.

The physical and mathematical analogy of this behavior of a set of resonances
to the super-radiance phenomenon was first indicated in \cite{SZPL88,SZNPA89}.
The starting point there was {\sl quantum chaos}, namely a question of the
influence of the continuum coupling on the statistical distribution of the
level spacings. It turns out that the super-radiant behavior {\sl survives} the
intrinsic dynamical chaos being mainly determined by the character of decay.
The spectral statistics, typical for canonical ensembles of random matrices,
evolve into the statistics of {\sl complex energies} which are at the same
time the poles of the scattering matrix in the space of open channels. The
statistical properties, in turn, are instrumental in defining the fluctuations
and correlations of cross sections of various reactions proceeding through
the channels open in the energy interval under study. The phenomenon of
super-radiance reveals itself in a specific two-cloud distribution of energies
and widths of resonances in the complex plane \cite{izr94,lehmann95}.

The adequate mathematical formalism for the description of this physics is
provided by the {\sl effective non-Hermitian Hamiltonian} \cite{SZAP92} suggested
in nuclear theory. The method goes back to Feshbach \cite{feshbach58,feshbach62}
and Fano \cite{fano61} extending and improving (eliminating not well defined
channel radii) the boundary condition method originated by Kapur and Peierls
\cite{kapur37}, Wigner and Eisenbud \cite{wigner47}, and Teichmann and Wigner
\cite{teichmann52}. Similar ideas can be found already in much earlier
notebooks by Majorana \cite{digrezia07}. The classical review articles are
available concerning this period of development
\cite{lane58,feshbachann58,breit59}. The equivalent method was independently
developed and applied, among other problems, to quantum scattering theory by
the outstanding Russian-Israeli mathematician M.S. Liv$\check{{\rm s}}$ic
\cite{livsic57,livsicbook}.

This approach was described in the well known book \cite{MW} although the fact
that the super-radiant physics directly follows from this Hamiltonian was
understood only later. The effective Hamiltonian, in general {\sl energy-dependent},
can be derived by a standard projection method when the part of the dynamics
corresponding to the continuum channels is formally eliminated and the whole
Schr\"{o}dinger equation is projected onto the internal dynamics. In this part
of Hilbert space the non-Hermiticity appears as a result of the presence of
open channels so that the eigenstates above threshold become unstable resonances,
and the eigenvalues of the Hamiltonian are {\sl complex energies} ${\cal E}=E-(i/2)\Gamma$,
where $\Gamma$ is the width of the resonance. The anti-Hermitian part of
the Hamiltonian has a specific factorized structure as follows from the unitarity
of the dynamics in the full space \cite{durand76}. In the regime of strong
continuum coupling this part plays the main role and the rank of the corresponding
matrix, equal to the number of open channels, determines the number of
super-radiant states. The Hermitian part of the effective Hamiltonian is
responsible for energy shifts of resonances and bound states; by its nature it
is similar to the Lamb shift but in this case of collective origin. It is useful
to stress that the non-Hermitian character of the Hamiltonian here is not related
to any approximation as long as all channels are accounted for.

In our review, the formalism of the effective Hamiltonian will serve as
a foundation of one of the most effective approaches to the description of
{\sl open many-body systems}. This is especially important for modern nuclear
physics where the center of interest moved to nuclei far from stability.
Such nuclei are often only marginally stable being close to the drip lines
which delineate the limits of existence for nuclear systems. The proximity
of the continuum $-$ both actual decay channels and virtual intermediate states $-$
determines the specific physics of such systems. The {\sl continuum shell model}
\cite{rotter91,VZcont03,VZCSM05,CSM06} developed in the framework of the effective
Hamiltonian allows one to describe in a unified approach bound states,
resonances and reaction cross sections. In distinction to other perspective
approaches to loosely bound nuclei, such as the {\sl Gamow shell model}, that
uses the analytical continuation to the complex plane \cite{michel03,michel09,garcia10},
here one can consider all processes at real energy of the reaction. The energy
dependence of the effective Hamiltonian reflects, in particular, the existence
of channel thresholds, and the proper behavior of the reaction amplitudes at
the thresholds is a noticeable and practically important feature of the
continuum shell model.

Nuclear physics is only one, even if essential, area of applications of the
ideas of the effective Hamiltonian and super-radiance. In fact, this is a
convenient way to describe dynamics of any quantum system in the energy region
including both bound states and continuum. Atomic autoionizing states
\cite{FGG96} and intermolecular processes \cite{verevkin88} were long ago
shown to reveal a similar behavior. Furthermore, we can expect the same
trends to manifest themselves in physics of intermediate and high energy
\cite{auerbach94,AZ02,AZV04}. The role of open channels in various
situations can also be played by the {\sl doorway states}
\cite{feshbachdoorway} which are forming a bottleneck connecting the states of
different nature \cite{auerbach07}; this is another generalization of the ideas of
super-radiance.

Important applications of the super-radiance and related theory can be found
in mesoscopic physics. The universal mesoscopic conductance fluctuations
\cite{beenakker97} share significant similarity with Ericson fluctuations
\cite{ericsonAP63,EMK66} typical for nuclear reactions observed in the energy
region of overlapping resonances \cite{weidenmueller90}. The detailed analysis
of correlations and fluctuations of cross sections for different channels can
shed new light on the fundamental problem of the signal transmission through a
quantum system, from simplest qubits to very complex molecules and artificial
structures with interfering propagation paths. The first results modifying
the Ericson theory by the effects of super-radiance and dependence of certain
observables on the degree of internal chaoticity appeared recently in the literature
\cite{celardo07,celardo08,sorathia09}, along with the applications to ordered
and disordered nanostructures \cite{celardo09,PRB10}. Vice versa, the intrinsic
chaos is strongly modified by the openness of the system.

Our review is devoted to the manifestations of the super-radiant mechanism outside
of quantum optics. The presentation is organized as follows. We start with illustrations of the
phenomenon of super-radiance in simple cases, Chapter 2. In Chapter 3 we
introduce the projection formalism, derive the effective Hamiltonian and show
how the generalized super-radiance emerges as a result of the strong coupling
of the intrinsic states to the continuum channels. As the first application of
the effective Hamiltonian we consider the nuclear continuum shell model.
Chapter 4 contains examples showing this approach at work. In Chapter 5 we show
the extension of the approach to doorway states and such collective excitations
as giant resonances. Other examples of the same approach, Chapter 6, will
illustrate applications to different areas of physics. We devote Chapter 7
to the problems of quantum chaos in open systems and their manifestations in
spectral statistics and reaction cross sections. Finally, Chapter 8 will discuss
the rapidly developing area of quantum signal transmission from the same
general viewpoint. The conclusion will be used mainly to list the interesting
problems which are still waiting for their explorer. We will try to explain the
main physical ideas along with their mathematical formulations but we will not
be able to go through detailed calculations referring instead to the original
literature.

\newpage

\section{Origin of super-radiance}

\subsection{Original idea}

For a simple start we will go back to the original idea of optical
super-radiance \cite{dicke54} and stress the elements which later will be
generalized to similar phenomena in various subfields of physics. Let us
consider an idealized system of $N$ identical two-level atoms confined to the
volume of a size much smaller than the wavelength of radiation that corresponds
to the transition between the levels of each qubit. We can describe the
intrinsic states $|\Psi\rangle$ of such a system with $N_{a}$ atoms in the
ground state and $N_{b}=N-N_{a}$ atoms in the excited state with the aid of
corresponding creation, $a^{\dagger}$ and $b^{\dagger}$, and annihilation, $a$
and $b$, operators as
\begin{equation}
|\Psi(N_{a},N_{b})\rangle\equiv |N_{a},N_{b}\rangle=
\frac{1}{\sqrt{N_{a}!N_{b}!}}\,(b^{\dagger})^{N_{b}}\,(a^{\dagger})^{N_{a}}|0\rangle,
                                                            \label{1}
\end{equation}
where for simplicity we assume that the atoms are bosons and $|0\rangle$
denotes the vacuum state.

Neglecting the radiation from the excited level, we have stable intrinsic space
spanned by the states (\ref{1}). It is useful to apply the Schwinger
representation of angular momentum \cite{schwinger} with the aid of
individual spins $s_{i}=1/2$ corresponding to each qubit and the total spin
${\bf S}=\sum_{i}{\bf s}_{i}$ of the entire system that is usually called here
the {\sl Bloch vector}. The identification of quantum numbers is obvious:
\begin{equation}
|\Psi(N_{a},N_{b})\rangle\Rightarrow |SM\rangle, \quad
S=\,\frac{1}{2}\,(N_{a}+N_{b})=\,\frac{1}{2}\,N, \quad
M=S_{z}=\,\frac{1}{2}\,(N_{b}-N_{a}),                         \label{2}
\end{equation}
or, in terms of original atomic operators,
\begin{equation}
|SM\rangle=[(S+M)!(S-M)!]^{-1/2}\,(b^{\dagger})^{S+M}(a^{\dagger})^{S-M}|0\rangle.
                                                         \label{3}
\end{equation}
The Hamiltonian of the closed system is trivial,
\begin{equation}
H=\omega\,\sum_{i}s_{iz}=\omega\,M,    \label{4}
\end{equation}
where the origin of the energy scale is set at the middle between the excited
and ground levels of an atom.

Now we make the excited atoms unstable introducing the radiation field. If
$A^{\ast}$ and $A$ are, in appropriate units, the amplitudes of radiation and
absorption of the photon, respectively, this part of the Hamiltonian is
\begin{equation}
H'=A^{\ast}a^{\dagger}b+Ab^{\dagger}a = A^{\ast}S_{-}+AS_{+}, \label{5}
\end{equation}
where, using the long wave limit, we neglect the phase differences between
the atoms. Each act of radiation brings $M$ by one step lower. If all atoms
are initially excited, $S=M=N/2$, the Bloch vector is at the north pole of
the Bloch sphere, and only spontaneous radiation of each atom is possible
with the total probability $|A^{\ast}\langle S,S-1|S_{-}|S,S\rangle|^{2}=
|A|^{2}2S=|A|^{2}N$. This is an incoherent process of independent radiation
by excited atoms with the intensity proportional to their number.

However, the same Hamiltonian (\ref{5}) describes also the processes of
interaction between the atoms though virtual radiation and absorption of
quanta. In this idealized model, such processes are amplified by the exact
degeneracy of atomic levels all tuned in resonance. While the spontaneous
radiation brings the average level of excitation down and the Bloch vector
deviates from the vertical position, the amplitude of radiation grows through
the virtual coupling between the radiators that becomes possible due to the
devastation of large-$M$ states and opening possibility of the virtual
absorption. The maximum is reached at $M=0$ (the Bloch vector in the equatorial
plane) being proportional to $|A|^{2}S(S+1)$. Here we have a coherent
spontaneous radiation $-$ {\sl super-radiance} $-$ with the probability
$\propto N^{2}$. Such a burst of spontaneous radiation was observed in
optically pumped HF \cite{skribanowitz73} and later studied in numerous
experiments. The underlying dynamics of the Bloch vector can be presented also
in a language of classical mechanics \cite{VZWNMP04}.

Of course, the practical implementation is not identical to this oversimplified
model. The different spatial positions of the atoms bring in the phase
differences between interfering radiators. (In fact, the seemingly opposite
limit of the size large compared to the wave length allows one to measure the
so-called {\sl collective Lamb shift} of the radiation line
\cite{rohlsberger10}; even a very small shift of this nature can limit accuracy
of atomic clocks \cite{beloy11}). The local nonuniformity and
collisional broadening lift the exact degeneracy of atomic levels. Direct,
for example dipole-dipole, interactions make the system different from the
collection of independent radiators, the extension of the model still allows
an exact solution \cite{dukelsky04,pan05}. However, even if not in such a pure form,
the super-radiance appears as an important physical effect with abundant
applications; the internal states of the radiators are effectively coupled
through the common radiation field.

For our goal, the main lesson is that the quantum states, along with possible
direct mixing by interactions, can be coupled also through the continuum of
open decay channels. Since the continuum coupling determines the width
$\Gamma$, or the lifetime $\tau\sim\hbar/\Gamma$, the states become
quasistationary and can be characterized by a {\sl complex energy},
\begin{equation}
{\cal E}=E-\,\frac{i}{2}\,\Gamma.                           \label{6}
\end{equation}
Similarly to standard perturbation theory, the efficiency of coupling is
determined by the ratio of the coupling strength to the energy spacing between
the coupled states. If the width $\Gamma$ is of the order of, or exceeds, the
spacing between real energies $E$, coupling through the continuum turns out to
be effectively strong. This transition to the regime of overlapping resonances
supplies the system by the new collectivity - the possibility of coherent
decay.

\subsection{Transport through a lattice}

The idea of generalized super-radiance works for many open quantum systems. We
illustrate this by a seemingly very different example \cite{SZAP92} taken from
condensed matter physics, where we will see the similar character of emerging
effects. Let us consider the simplest one-dimensional crystal with identical
cells numbered $1,2,...,N$. A particle has levels at energy $\epsilon$ in all
cells and the hopping amplitude $v$ to adjacent cells. The edge barriers allow
the coupling to environment so that the corresponding decay widths are
$\gamma_{L}$ and $\gamma_{R}$. The system can be described by an effective
non-Hermitian Hamiltonian with matrix elements in the site representation
\begin{equation}
{\cal H}_{nm}=\epsilon\delta_{nm}+v\Bigl(\delta_{m,n+1}+\delta_{m,n-1}\Bigr)-
\frac{i}{2}\,\Bigl(\gamma_{L}\delta_{n1}\delta_{m1}+\gamma_{R}\delta_{nN}\delta_{mN}\Bigr)
                                                           \label{7}
\end{equation}
which describe the decay to outside with the help of complex energies (\ref{6})
at the edge sites. In the absence of decay, the internal levels would be split
by the hopping and form a crystal band of Bloch waves labeled by the discrete
wave vectors $q=1,...,N$. The band is characterized by corresponding energies
(we set $\epsilon=0$)
\begin{equation}
\epsilon_{q}= 2v\cos\varphi_{q}, \quad \varphi_{q}=\,\frac{\pi}{N+1}\,q,  \label{8}
\end{equation}
and the eigenvectors $|q\rangle$ with components in the site representation
\begin{equation}
\langle n|q\rangle=\sqrt{\frac{2}{N+1}}\,\sin(n\varphi_{q}).     \label{9}
\end{equation}
In this basis, the Hamiltonian (\ref{7}) is transformed to
\begin{equation}
{\cal H}_{qq'}=\epsilon_{q}\delta_{qq'}-\,\frac{i}{2}\,W_{qq'},       \label{10}
\end{equation}
with the factorized matrix elements of the anti-Hermitian part,
\begin{equation}
W_{qq'}=\,\frac{2}{N+1}\,\sin\varphi_{q}\,\sin\varphi_{q'}[\gamma_{L}
+(-)^{q+q'}\gamma_{R}].                                       \label{11}
\end{equation}
As the eigenstates are delocalized, each of them acquires a width; the maximum
width appears in the middle of the energy band. The intrinsic hopping interaction
is now substituted with the anti-Hermitian interaction through the continuum.
The quasistationary states can be found by diagonalizing the effective
Hamiltonian (11). The complex energies (\ref{6}) are given by the roots of
the secular equation
\begin{equation}
1+i(\gamma_{L}+\gamma_{R})P_{+}({\cal E})+\gamma_{L}\gamma_{R}[P_{-}^{2}({\cal E})
-P_{+}^{2}({\cal E})]=0,                               \label{12}
\end{equation}
where
\begin{equation}
P_{\pm}({\cal E})=\frac{1}{N+1}\,\sum_{q}(\pm)^{q}\,\frac{\sin^{2}\varphi_{q}}
{{\cal E}-\epsilon_{q}}.                                      \label{13}
\end{equation}

In all cases the total width of all resonances $\sum_{q}\Gamma_{q}$ has to be equal
to the trace, $\gamma_{L}+\gamma_{R}$, of $W$. Two possible limiting situations
clearly come out. At weak continuum coupling, the individual Bloch states give rise
to isolated resonances which can still be labeled by the wave vector $q$. Neglecting
the last term in eq. (\ref{12}), we find the widths as the diagonal elements of $W$,
\begin{equation}
{\cal E}_{q}\approx \epsilon_{q}-\,\frac{i}{2}\Gamma_{q}, \quad
\Gamma_{q}=W_{qq},                                    \label{14}
\end{equation}
which, by use of eq. (\ref{9}), can be presented in an obvious form
\begin{equation}
\Gamma_{q}=\gamma_{L}|\langle 1|q\rangle|^{2}+\gamma_{R}|\langle N|q|\rangle|^{2}.
                                                                    \label{15}
\end{equation}
The decay here serves as an analyzer singling out the components of the quasistationary state
matched to the specific decay channel, left or right in this model. Such arguments
are usually used as a justification \cite{brody81} for identifying the width distribution
in a compound nucleus with the distribution of the single components of the compound
nuclear wave function. For reduced neutron widths of low-lying neutron resonances with
Gaussian distribution of basis components this gives rise to the {\sl Porter-Thomas
distribution} (PTD) \cite{PT56}. However, this result and underlying logic are valid
only in the lowest approximation with respect to {\sl weak continuum coupling} of
non-overlapping unstable states.

In the opposite limiting case of {\sl strong continuum coupling}, the lifetime with respect
to the irreversible decay outside is shorter than for the hopping between the sites,
$\gamma_{L,R}\gg v$. In this limit (and $N\gg 1$), one can neglect the band splitting
$\epsilon_{q}$ in the denominators of $P_{\pm}$ in eq. (\ref{13}), so that $P_{-}$
vanishes while $P_{+}\rightarrow 1/(2{\cal E})$. Then two roots are shifted along the
imaginary energy axis defining
\begin{equation}
{\cal E}_{L,R}=-\,\frac{i}{2}\,\gamma_{L,R}.                    \label{16}
\end{equation}
This limit of short-lived collective states is an analog of optical Dicke
super-radiance. (If $\gamma_{L}$ and $\gamma_{R}$ significantly differ, the
system will reveal two super-radiant transitions \cite{PRB10}). In
distinction to usual collective states in a many-body system, here the
collectivization proceeds through the widths of resonances, all Bloch waves
radiate coherently being coupled through the decay amplitudes. In the
$q$-space, the super-radiant states are in the middle of the spectrum, while in
the site representation it is trivial to find that they are located at the
edges. Therefore, at strong continuum coupling it might be useful to go to the
{\sl doorway representation} \cite{SZAP92} that will be discussed in the corresponding
section from a more general viewpoint. In the one-channel case, the transformation
of the $S$-matrix to such a doorway form can be made explicitly as was shown long
ago \cite{jeukenne69}. In the limit (\ref{16}), remaining $N-2$ intrinsic states are
{\sl dark} being decoupled from the continuum. It is easy to find the correction to
eq. (\ref{16}), where the long-lived (trapped) states acquire small widths,
$\propto (v/\gamma_{L,R})^{2}$, through the intermediate coupling to the doorway
states directly coupled to the continuum. We can also mention that the disorder
in the closed system may even amplify the effects of the super-radiance
\cite{VZWNMP04} through the formation of localized states more easily decaying
outside. We return to the more detailed discussion of quantum transport in Sec. 8.

The common features of our two examples - optical super-radiance and signal transmission
through a chain of radiators - are generic for all problems with open quantum systems.
We have here a kind of phase transition from independent unstable states to collective
radiation (or any process developing with, or through, open continuum channels).
A typical control parameter regulating this transformation is the relative strength
of continuum coupling with respect to the level spacing in the closed system.
Below we introduce the universal formalism adjusted for the consistent description
of such situations.

\newpage

\section{Open quantum system}

\subsection{Projection formalism}

Following \cite{feshbach58,feshbach62,MW} we consider a general quantum system
that has both bound and continuum states. The states $|c;E\rangle$ in the
continuum will be labelled by total energy $E$ and the quantum numbers of the
{\sl channels}. The channel, $c$, is characterized by the kind(s) of the
fragments in the continuum, their intrinsic states, and all discrete quantum
numbers (spins, isospins, parity...). Every channel $c$ opens at its
{\sl threshold energy} $E^{c}$. The whole dynamics in the Hilbert space
of the system is governed by the Hermitian Hamiltonian $H=H^{\dagger}$, while
the reactions $a\rightarrow b$ between various channels, including the elastic
one, $b=a$, can be described by the scattering matrix in the channel space,
$S^{ba}(E)$, which is {\sl unitary}, $S^{\dagger}=S^{-1}$.

Using the stationary quantum mechanics we have to solve the Schr\"{o}dinger
equation for the wave function at real energy,
\begin{equation}
H\Psi_{E}=E\Psi_{E},                                    \label{17}
\end{equation}
where energy $E$ can belong to the discrete spectrum (and then has to be
determined by the solution of the equation) or to the continuum, if some
channels are {\sl open} having their thresholds below given value of $E$. In
the coordinate representation, those two cases have different boundary
conditions. The threshold positions are also to be determined self-consistently
by the Schr\"{o}dinger equation.

It is always possible to formally subdivide the Hilbert space of the problem into
two (or more, see for example \cite{karataglidis08}) complementary parts using
the expansion of the unit operator into arbitrarily selected orthogonal projectors,
\begin{equation}
\hat{{\sl 1}}={\cal P}+{\cal Q}, \quad {\cal PP}={\cal P}, \quad {\cal QQ}={\cal
Q}, \quad {\cal PQ}={\cal QP} = 0.                        \label{18}
\end{equation}
Then the Hamiltonian $H$, as any other operator, contains four classes of
matrix elements,
\begin{equation}
H=H_{{\cal Q}{\cal Q}}+H_{{\cal Q}{\cal P}}+H_{{\cal P}{\cal Q}}+H_{{\cal
P}{\cal P}},                                             \label{19}
\end{equation}
while the wave function has two parts, $\Psi={\cal P}\Psi+{\cal Q}\Psi$, which
satisfy the coupled set of equations,
\begin{equation}
[H_{{\cal Q}{\cal Q}}-E]{\cal Q}\Psi=-H_{{\cal Q}{\cal P}}{\cal P}\Psi, \quad
[H_{{\cal P}{\cal P}}-E]{\cal P}\Psi=-H_{{\cal P}{\cal Q}}{\cal Q}\Psi.
                                                            \label{20}
\end{equation}
Formally eliminating the part ${\cal P}\Psi$ of the wave function,
\begin{equation}
{\cal P}\Psi=\Bigl(E-H_{{\cal P}{\cal P}}\Bigr)^{-1}\,H_{{\cal P}{\cal Q}}{\cal
Q}\Psi,                                                  \label{21}
\end{equation}
we find the equation projected into the class ${\cal Q}$,
\begin{equation}
{\cal H}(E){\cal Q}\Psi=E{\cal Q}\Psi,                   \label{22}
\end{equation}
where we have to deal with the {\sl effective energy-dependent Hamiltonian}
\begin{equation}
{\cal H}(E)=H_{{\cal QQ}}+H_{{\cal QP}}\,\frac{1}{E-H_{{\cal PP}}}\,H_{{\cal
PQ}}.                                                     \label{23}
\end{equation}
It is useful to stress that the formal results here and below do not require
any explicit expressions for the projection operators although it can be done
in many ways \cite{feshbach62}, both in Hilbert space and in configuration
space. The flexibility of the approach is its important advantage. Working in
the Hilbert space we can avoid problems related to the dependence on the
coordinate radius of a channel and explicit matching of the solutions at such a
surface \cite{thomas55,bloch57,descouvemont10}.

The formal solution (\ref{21}) has to be complemented with rules of working
with the singularities emerging when $H_{{\cal PP}}$ has eigenvalues at energy
$E$. Although the projection procedure can use any subdivision of space, we
will assume that the part ${\cal P}$ may contain the continuous spectrum
$|c;E\rangle$, so that the second part of eq. (\ref{20}) has a solution of
the homogeneous equation, and in any channel $c$ the singularity appears at
energy above the corresponding threshold ($E\geq E^{c}$). The standard rule will be applied
that real energy $E$ is to be considered as a limiting value from the upper
half of the complex energy plane, $E^{(+)}=E+i0$. With this rule, those states
of space ${\cal Q}$ which will turn out to be coupled to open channels in
${\cal P}$, will acquire the outgoing waves and become unstable. Then the
matrix elements of the effective Hamiltonian (\ref{23}) can be written more in
detail as
\[{\cal H}_{12}=H_{12}+\sum_{c'c''}\int (d\tau'\,d\tau'')dE'dE''\]
\begin{equation}
\times \langle 1|H_{{\cal QP}}
|c',\tau',E'\rangle\left(\frac{1}{E^{(+)}-H_{{\cal
PP}}}\right)_{c',\tau',E';c'',\tau'',E''} \langle c'',\tau'',E''|H_{{\cal
PQ}}|2\rangle.                                                  \label{24}
\end{equation}
Here we use an arbitrary orthonormal basis in the ${\cal Q}$-space with the basis
states labeled as $|1\rangle,\,|2\rangle,...$ The differentials $(d\tau'\,d\tau'')$
symbolically denote kinematic variables in the channels $c',c''$, apart from
explicitly indicated energy.

It is always possible \cite{engelbrecht73} to find the {\sl eigenchannel} basis
in the ${\cal P}$-space that brings $H_{{\cal PP}}$ to the diagonal form with
eigenvalues $E'(c;\tau)$ and the Hamiltonian (\ref{24}) simplifies to
\begin{equation}
{\cal H}_{12}=H_{12}+\sum_{c}\int (d\tau')dE'\langle
1|H_{\cal{QP}}|c,\tau',E'\rangle \,\frac{1}{E^{(+)}-E'(c,\tau')}\langle
c,\tau',E'|H_{{\cal PQ}}|2\rangle. \label{25}
\end{equation}
For channels closed at energy $E$, there is no singularity, and the superscript
$(+)$ can be omitted. For open channels we use the usual identity valid for
real $x$,
\begin{equation}
\frac{1}{E^{(+)}-x}=\,{\tt P.v.}\,\frac{1}{E-x}\,-i\pi\delta(E-x),  \label{26}
\end{equation}
where {\tt P.v.} is a symbol of the principal value. This leads us to the
``canonical" form of the effective Hamiltonian,
\begin{equation}
{\cal H}=\widetilde{H}-\,\frac{i}{2}\,W.                          \label{27}
\end{equation}
The Hermitian part here is
\begin{equation}
\widetilde{H}=H+\Delta(E),                                       \label{28}
\end{equation}
where $\Delta(E)$ is an {\sl off-shell} integral in the sense of the principal
value,
\begin{equation}
\Delta_{12}(E)=\,{\tt P.v.}\sum_{c}\int (d\tau')dE'\langle
1|H_{\cal{QP}}|c,\tau',E' \rangle \,\frac{1}{E-E'(c,\tau')}\langle
c,\tau',E'|H_{{\cal PQ}}|2\rangle,
                                                             \label{29}
\end{equation}
while the anti-Hermitian part is given by
\begin{equation}
W_{12}(E)=2\pi \sum_{c({\rm open})}A_{1}^{c}(E)A_{2}^{c\ast}(E),  \label{30}
\end{equation}
and we introduced the effective amplitudes $A_{1}^{c}(E)$ as matrix elements of
$H_{{\cal QP}}$ taken {\sl on-shell}, $E'=E$, with all kinematic factors coming
from the integral included into their definition. It is important to notice
that the real ({\sl dispersive}) part $\Delta$ of the effective Hamiltonian,
sometimes called the {\sl energy shift}, includes the sum over {\sl all channels},
open and closed, while the imaginary ({\sl absorptive}) part $W$ contains only
contributions of the channels {\sl open at given energy}. The kinematic factors
(the density of states in the continuum) guarantee that any amplitude
$A_{1}^{c}(E)$ vanishes at the threshold energy $E^{c}$. By construction,
these amplitudes are combined into a factorized structure that is a consequence
(and a mathematical reason) of the unitarity of the scattering matrix
\cite{durand76}. We have already seen this factorization in a specific
simple model, eq. (\ref{11}).

\subsection{Effective Hamiltonian and reaction theory}

The effective Hamiltonian (\ref{27}) allows us to consider the properties of an
open system in a very general fashion. The advantage of this formulation is in
the discretization of the problem since we can assume that the internal part
of the Hamiltonian, $H_{{\cal QQ}}$, by construction of our subdivision (\ref{19})
has only a discrete spectrum, so that the resonances can appear only due to the
continuum coupling, $H_{{\cal PQ}}$ and $H_{{\cal QP}}$. This ``natural"
discretization should not be confused with the discretization of the continuum
used frequently in the numerical studies of various reactions \cite{CDCC}; here
we are still working within the exact formulation of the theory but the whole
dynamics is projected onto the ``intrinsic" part of the total Hilbert space.

For a system with $N$ as a (usually very large) dimension of the intrinsic
space and $M$ open channels, the amplitudes $A_{1}^{c}$ can be considered
as rectangular matrices $N\times M$ which we can label ${\bf A}$, so that
the matrix of the operator $W$, eq. (\ref{30}), can be presented as a product
\begin{equation}
W=2\pi\,{\bf A}{\bf A}^{\dagger}.                              \label{31}
\end{equation}
In a time-reversal invariant system, the amplitudes $A_{1}^{c}$ can be taken
as real numbers; then ${\bf A}^{\dagger}={\bf A}^{T}$. This factorized
structure facilitates the task of establishing the relation of this approach to
the formal theory of reactions \cite{wu62,goldberger64}. On the other hand, this
is a mathematical key to super-radiance.

We can introduce the propagator at complex energy ${\cal E}$ for the effective
Hamiltonian ${\cal H}$ as
\begin{equation}
{\cal G}({\cal E})=\,\frac{1}{{\cal E}-{\cal H}},          \label{32}
\end{equation}
where physical energy axis is a limit ${\cal E}\rightarrow E^{(+)}$. The analog
of the Dyson equation that connects ${\cal G}$ with the propagator $G$ for a
closed system,
\begin{equation}
G({\cal E})=\,\frac{1}{{\cal E}-\tilde{H}},                 \label{33}
\end{equation}
can be derived algebraically with the help of the factorized structure of $W$:
\begin{equation}
{\cal G}({\cal E})=G({\cal E})-\,\frac{i}{2}\,G({\cal E}){\bf A} \,\frac{1}
{1+(i/2)\hat{K}({\cal E})}\,{\bf A}^{\dagger}G({\cal E}).     \label{34}
\end{equation}
Here and later the hats mark $M\times M$ matrices in the channel
space; in particular, the matrix
\begin{equation}
\hat{K}({\cal E})={\bf A}^{\dagger}G({\cal E}){\bf A}            \label{35}
\end{equation}
describes the full propagation inside the system from the entrance channel to
the exit channel. The series of such propagations summed over in the operator
$[1+(i/2)\hat{K}]^{-1}$ of eq. (\ref{34}) corresponds to the full process of
the reaction through the system by an infinite number of intermediate
excursions from the intrinsic space to the continuum. The matrix (\ref{35})
substitutes in this theory the standard $R$-matrix \cite{lane58,breit59,descouvemont10}.

The part of the amplitude for the reaction $b\rightarrow a$ that proceeds
through the states of the subspace ${\cal Q}$ is now an element $(T_{{\cal
Q}})^{ab}$ of the $M\times M$ matrix
\begin{equation}
\hat{T}_{{\cal Q}}({\cal E})={\bf A}^{\dagger}{\cal G}({\cal E}){\bf A}=
\hat{K}({\cal E})\,\frac{1}{1+(i/2)\hat{K}({\cal E})}.     \label{36}
\end{equation}
The total reaction amplitude includes also the {\sl direct processes}
determined by the dynamics in the ${\cal P}$-space without entering the
intrinsic state,
\begin{equation}
\hat{T}=\hat{T}_{{\cal P}}+\hat{T}_{{\cal Q}},              \label{37}
\end{equation}
where ${\cal E}\rightarrow E^{(+)}$. The item $\hat{T}_{{\cal P}}$ is
responsible for what is called {\sl potential scattering}. The complete {\sl
scattering matrix} can be written as
\begin{equation}
\hat{S}({\cal E}) =1-i\hat{T}({\cal E})=-i\hat{T}_{{\cal P}}+\frac{1-(i/2)
\hat{K}({\cal E})}{1+(i/2)\hat{K}({\cal E})},               \label{38}
\end{equation}
and the cross sections in open channels are defined by the corresponding matrix
elements of the $\hat{S}$-matrix {\sl at physical energy}, with the interference of
direct and resonance scattering taken into account. The amplitudes ${\bf A}$ also
contain phases of the potential scattering at the entrance and exit of the channels.
The resonance part, closely related to the intrinsic structure of the system, will
capture our main attention. In the absence of direct processes, this part is evidently
unitary by itself.

The amplitudes $T^{ab}_{{\cal Q}}({\cal E})$, according to the first equality
(\ref{36}), have the poles in the complex plane of energy ${\cal E}$ coming
from the poles of ${\cal G}({\cal E})$, which are the eigenvalues (\ref{6}) of
the effective Hamiltonian ${\cal H}$. Those eigenvalues determine the
``compound-nucleus" resonance spectrum of the system. We use here this term
only to express the fact that the wave function of such a state has components in
the ${\cal Q}$-space. Whether a resonance indeed agrees with the notion of
compound nucleus in the standard meaning of this concept can be decided by the
comparison of the resonance lifetime $\tau\sim\hbar/\Gamma$ with the
characteristic equilibration time inside the system. As stated by Feshbach
\cite{feshbach62}, ``the effective Hamiltonian method is most convenient for
nuclear reactions in that it differentiates ... between compound and direct
nuclear reactions". On the real energy axis, the amplitudes have branch points
at the channel thresholds.

\subsection{Emergence of resonances}

We start with a simple example of a resonance that emerges as a result of
interaction between the two subspaces (\ref{18}). This will illustrate a direct
relation between the method under discussion and the textbook scattering
theory. The formalism, apparently first used by Fano \cite{fano61} in
application to the autoionizing states of an atom, immediately follows from the
general projection method. In fact, this approach is similar to the
consideration of the spreading width of a collective (``{\sl bright}") state
interacting with the background of secondary states and in such a form it was
suggested by Rice \cite{rice33}. This analogy will reappear in the discussion
of the doorway states in Sec. 6.

If we consider only a small number of special intrinsic states, it is convenient
to project the entire dynamics into the external ${\cal P}$-space. Eliminating
the part ${\cal Q}\Psi$ of the wave function, we come to the equation
\begin{equation}
[H_{{\cal PP}}-E]{\cal P}\Psi=H_{{\cal PQ}}\,\frac{1}{H_{{\cal QQ}}-E}\,
H_{{\cal QP}}{\cal P}\Psi.                                  \label{39}
\end{equation}
Here energy $E$ belongs to the continuous spectrum, and the homogeneous
equation, $[H_{{\cal PP}}-E']{\cal P}\Psi=0$, has solutions $\psi^{c}_{E'}$
corresponding to the asymptotic channels $c$ included in the ${\cal P}$-space.
The right hand side of eq. (\ref{39}) describes the intrinsic states with the
same conserved quantum numbers which are coupled to the continuum and, if above
thresholds, can reveal autoionization. As a demonstration of principles,  we
consider here only the case of one intrinsic state $\Phi$ with unperturbed
energy $H_{\cal{QQ}}=\epsilon$ coupled to a single channel [if not all open
channels are included into the ${\cal P}$-space, we have to consider the limit
$E\rightarrow E^{(+)}$ in the denominator of the second term in eq.
(\ref{39})].

The solution of eq. (\ref{17}) is a superposition
\begin{equation}
\Psi_{E}=a(E)\Phi+\int dE'\,b(E,E')\psi_{E'},          \label{40}
\end{equation}
a direct analog of the spreading of a bright state in the discrete spectrum.
Omitting from our notations the dependence of $a$ and $b$ on running energy
$E$, we obtain the set of coupled equations for the amplitudes $a$ and $b(E')$
in terms of the continuum coupling amplitudes $A(E')=\langle\Phi|H_{{\cal QP}}|
\psi_{E'}\rangle$, as in eq. (\ref{30}),
\begin{equation}
(E-\epsilon)a=\int dE'\,A(E')b(E'), \quad (E-E')b(E')=A^{\ast}(E')a. \label{41}
\end{equation}
The existence of the homogeneous solution in the continuum at $E'=E$ leads to
the general result in the form of the principal value plus the delta-function,
\begin{equation}
b(E')=\Bigl[{\tt P.v.}\frac{1}{E-E'}\,+r(E)\delta(E-E')\Bigr]A^{\ast}(E')a,
                                                                  \label{42}
\end{equation}
where the residue factor $r(E)$ is {\sl real} and has to be determined
self-consistently, while in the analogous situation in eq. (\ref{25}) we could
simply use the standard identity (\ref{26}). The first equation of (\ref{41})
determines now
\begin{equation}
r(E)=\,\frac{E-\epsilon-\Delta(E)}{|A(E)|^{2}},                \label{43}
\end{equation}
where $\Delta(E)$ is the simplified form of the principal value integral
(\ref{29}). Introducing the coordinate representation of the scattering
problem, one can show \cite{fano61} that this residue defines the phase shift
$\delta$,
\begin{equation}
\tan\delta(E)==-\,\frac{\pi}{r(E)}.                              \label{44}
\end{equation}

Finally, the normalization of the full wave function by the delta-function of
energies in the continuum, allows one to find the remaining amplitude
\begin{equation}
|a(E)|^{2}=\,\frac{|A(E)|^{2}}{[E-\epsilon-\Delta(E)]^{2}+\Gamma^{2}/4},
                                                                 \label{45}
\end{equation}
of resonance shape with the energy shift $\Delta(E)$ and the width
\begin{equation}
\Gamma(E)=2\pi|A(E)|^{2},                                      \label{46}
\end{equation}
in agreement with the definition of eq. (\ref{30}). Both, the energy shift and
the width, are {\sl energy-dependent}. The shift $\Delta(E)$ vanishes if one
can neglect the energy variation of the amplitudes $A(E)$. The physical
characteristics changing rapidly as a function of energy near the resonance
peak lead to the highly asymmetric curves for cross sections which is a well
known feature of autoionization; the typical example of the $2s2p$ excited
state in the helium atom was analyzed in detail in the original work
\cite{fano61}.

\subsection{Overlapping resonances and super-radiance}

Our next step is the analysis of the eigenvalues and the eigenfunctions of the
effective non-Hermitian Hamiltonian ${\cal H}$. To emphasize the physics of
super-radiance and related transformations of the resonance spectrum, we
neglect for a while the energy dependence of the amplitudes $A_{1}^{c}$. This
assumption will be lifted in the continuum shell model when the threshold
dependence of the amplitudes will be important for the spectroscopy of the
resonances. If our energy is not close to thresholds, the energy dependence is
smooth. In the case of many resonances with the same quantum numbers, such as
the low-energy neutron resonances in heavy nuclei, the energy dependence is
common for all of them and can be usually subtracted \cite{brody81} using
the so-called {\sl reduced widths}. Special precaution should be taken if
the compound resonances come from the single-particle resonance that served
as a doorway state on the way to a compound nucleus \cite{weidenmueller10}.
In this case the shape of the doorway strength function may distort the reduced
widths (we return to the width distribution in Sec. 7.2).

The two parts of the effective Hamiltonian (\ref{27}) do not commute in general
and cannot be brought simultaneously to a diagonal form by an orthogonal
transformation of the basis. The complex diagonalization of ${\cal H}$ leads to
a {\sl biorthogonal} set $\{\chi_{r},\chi^{\ast}_{r}\}$ of right and left eigenfunctions
which have complex conjugate eigenvalues ${\cal E}_{r}$ and ${\cal E}^{\ast}_{r}$.
All eigenvalues are located in the lower part of the complex energy plane; the
resonance widths $\Gamma_{r}$, eq. (\ref{6}), are positive. This can be easily
seen from the properties of the effective Hamiltonian (\ref{27}). Its
eigenvalue equation reads
\begin{equation}
\Bigl[{\cal E}_{r}-H_{{\cal QQ}}-H_{{\cal QP}}\,\frac{1}{E^{(+)}-H_{{\cal PP}}}
\,H_{{\cal PQ}}\Bigr]\chi_{r}=0.                     \label{47}
\end{equation}
Taking the matrix element with the biorthogonal partner $\chi^{\ast}_{r}$, we
obtain
\begin{equation}
{\cal E}_{r}-{\cal E}^{\ast}_{r}=\left\langle\chi_{r}^{\ast}\left|H_{{\cal QP}}
\left(\frac{1}{E^{(+)}-H_{{\cal PP}}}\,-\,\frac{1}{E^{(-)}-H_{{\cal PP}}}\right)
H_{{\cal PQ}}\right|\chi_{r}\right\rangle.                                \label{48}
\end{equation}
The principal value terms cancel, and the imaginary part of the eigenenergy
determines the positively defined width (the so-called {\sl Bell-Steinberger
relation}):
\begin{equation}
\Gamma_{r}=2\pi\langle\chi_{r}^{\ast}|H_{\cal{QP}}\delta(E-H_{{\cal PP}})
H_{{\cal PQ}}|\chi_{r}\rangle.                                   \label{49}
\end{equation}

It might be convenient to work instead with a complete orthogonal basis of one
of the parts, $\widetilde{H}$ or $W$, which will be distinguished as the {\sl
internal} and {\sl doorway representation}, respectively. Let $|\alpha\rangle$
be the eigenfunctions of $\widetilde{H}$ with real eigenvalues [as the Bloch waves
in the case of the lattice, eqs. (\ref{8}) and (\ref{9})],
\begin{equation}
\widetilde{H}|\alpha\rangle=\epsilon_{\alpha}|\alpha\rangle.       \label{50}
\end{equation}
The part $W$ of the total Hamiltonian remains factorized after the unitary
transformation to the basis $\alpha\rangle$, compare eq. (\ref{11}),
\begin{equation}
W_{\alpha\beta}=2\pi\sum_{1,2;c}\langle\alpha|1\rangle A_{1}^{c}A_{2}^{c\ast}
\langle 2|\beta\rangle\equiv 2\pi\sum_{c}B_{\alpha}^{c}B_{\beta}^{c\ast}.
                                                              \label{51}
\end{equation}
In the internal representation, we deal with the Hamiltonian
\begin{equation}
{\cal H}_{\alpha\beta}=\epsilon_{\alpha}\delta_{\alpha\beta}-\,\frac{i}{2}\,
2\pi\sum_{c({\rm open})}B_{\alpha}^{c}B_{\beta}^{c\ast}.         \label{52}
\end{equation}
According to eq. (\ref{34}), the complex eigenvalues are the roots of
\begin{equation}
{\rm Det}\,\left\{1+\,\frac{i}{2}\,\hat{K}({\cal E})\right\}=0. \label{53}
\end{equation}

Two opposite limiting regimes naturally come into consideration. At {\sl weak coupling}
to continuum, the anti-Hermitian part is a perturbation to a standard intrinsic
Hamiltonian of the shell-model type. Then we expect the appearance of narrow
resonances corresponding to long-lived states barely coupled to decay channels.
At {\sl strong coupling}, the main role is played by the part $W$. In general,
the whole picture including the resonances depends on the running energy $E$ of
the system, although this might be not important far from thresholds.

Let us make simple estimates assuming that the intrinsic states
$|\alpha\rangle$ are approximately of the same degree of complexity and have a
similar strength of continuum coupling so that it makes sense to use the
average parameters. The transformation (\ref{51}) preserves the trace of $W$
which we denote $w$. This trace will be the same after the final complex
transformation to the eigenbasis of ${\cal H}$, and the typical width of a
resonance in the case of weak coupling will be $\Gamma\sim w/N$. If this width
is small compared to the spacings $D$ between the levels $\epsilon_{\alpha}$,
the usual perturbation theory should be valid. This can be expressed with the
aid of the coupling parameter $\kappa\sim(\Gamma/D)\ll 1$, or, in global
characteristics, $w/(ND)\ll 1$, the total summed width is small compared to the
energy interval covered by the relevant intrinsic levels. Only the diagonal
part of $W$ in the representation (\ref{52}) is important in this limit, and
the resonance states, as in the example (\ref{15}), still can be labeled by
the quantum numbers of $\widetilde{H}$, eq. (\ref{50}),
\begin{equation}
{\cal E}_{r}\Rightarrow {\cal E}_{\alpha}=\epsilon_{\alpha}-\,\frac{i}{2}\,
\Gamma_{\alpha}, \quad \Gamma_{\alpha}=2\pi\sum_{c}|B^{c}_{\alpha}|^{2}.
                                                             \label{54}
\end{equation}
In this regime the reactions reveal narrow isolated resonances $|\alpha\rangle$
with {\sl partial widths}
\begin{equation}
\gamma_{\alpha}^{c}=2\pi|B^{c}_{\alpha}|^{2}, \quad \Gamma_{\alpha}=\sum_{c}
\gamma_{\alpha}^{c}.                                          \label{55}
\end{equation}
This is, for example, the typical picture for low-energy neutron resonances.

The situation drastically changes when we approach the strong coupling regime,
$\kappa\sim 1$. Then $w\sim ND$, the resonances start to overlap and the
situation of similar isolated resonances turns out to be unstable with
respect to the sharp redistribution of widths. To illustrate this point we
start with the {\sl one-channel} case, when eq. (\ref{53}) reads
\begin{equation}
1+\,\frac{i}{2}\,\sum_{\alpha}\,\frac{\gamma_{\alpha}}{{\cal E}
-\epsilon_{\alpha}}=0.                                   \label{56}
\end{equation}
In the same basis we now can see a very different solution in the limit of
$\kappa\gg 1$, namely an emergence of a very broad state that accumulates
almost the whole summed width $w=\sum_{\alpha}\gamma_{\alpha}$. If $w\gg ND$,
all intrinsic states would be covered by the total width, so that introducing
the centroid of real energies weighted with their original widths
$\gamma_{\alpha}$,
\begin{equation}
\bar{\epsilon}=\sum_{\alpha}\epsilon_{\alpha}\, \frac{\gamma_{\alpha}}{w},
                                                        \label{57}
\end{equation}
we immediately find from eq. (\ref{56}) the broad pole ${\cal E}_{1}=E_{1}-
(i/2)\Gamma_{1}$ with ${\cal E}_{1}$ close to the centroid (\ref{57}) and the
width $\Gamma_{1}$ close to $w$. The remaining $N-1$ states are in this limit very narrow
(trapped). With corrections of the second order in $1/\kappa\ll 1$, this
special root of eq. (\ref{56}) is given by
\begin{equation}
E_{1}=\bar{\epsilon}+4\sum_{\alpha}\epsilon_{\alpha}\,\frac{(\epsilon_{\alpha}-
\bar{\epsilon})^{2}\gamma_{\alpha}}{w^{3}},               \label{58}
\end{equation}
\begin{equation}
\Gamma_{1}=w-4\sum_{\alpha}\,\frac{\gamma_{\alpha}(\epsilon_{\alpha}
-\bar{\epsilon})^{2}} {w^{2}}.                        \label{59}
\end{equation}
The trapped states have narrow widths $\Gamma_{t}\sim w/(N\kappa^{2})$ which
are much smaller than the average starting width $\gamma\sim w/N$. The more detailed
analytical solutions can be derived in a model \cite{FGG96} with equidistant
intrinsic spectrum and weakly fluctuating $\gamma_{\alpha}$.

This example shows a generic mechanism of the formation of the super-radiant
state. When the widths of individual resonances overlap and $\kappa>1$, the
dominating interaction between the resonances proceeds through the common
channels in the continuum. Roughly speaking, energy uncertainty of an unstable
state makes a virtual decay of this state possibly end with the return in the
${\cal Q}$-space to a neighboring overlapped state. This coupling leads to the
resonance collectivization and the segregation of a super-radiant combination,
the story one can identify in the two examples discussed in Sec. 2. It is easy
to see that the super-radiant state $|\Phi_{1}\rangle$ is a superposition of
intrinsic states best adjusted to the decay channel. Expressed as a vector in
${\cal Q}$-space, it is nearly aligned along the decay amplitude vector ${\bf
B}=\{B_{\alpha}\}$, its transverse components are small, $\sim 1/\kappa$. In
the example of a periodic lattice of Sec. 2, this corresponded to the
localization of the super-radiant states at the spatial edges of the chain
opened for the decay. The trapped states return to the non-overlap regime and
belong mainly to the subspace orthogonal to ${\bf B}$.

We can also establish the hierarchy of time intervals corresponding to various
parts of the dynamics involved in this transition. At $\kappa \gg 1$, the longest
time is the lifetime of the trapped states, $\tau_{t}\sim\hbar/\Gamma_{t}\sim
\kappa(\hbar/D)\gg \hbar/D$, where the Weisskopf time $\tau_{W}\sim\hbar/D$
\cite{weisskopf37} describes the recurrence of the intrinsic wave packet in its
evolution in the closed space without decays. The inequality $\tau_{t}>\tau_{W}$
means that the wave functions of trapped states can be considered fully
equilibrated as corresponds to the proper compound stage. The next time $\hbar/ND$ is
even shorter than the recurrence time representing essentially the lifetime of
a certain intrinsic basis state $|1\rangle$ with respect to the intrinsic
spreading. Finally, the lifetime of the super-radiant state is the shortest,
$\tau_{s}\sim\hbar/\Gamma_{1}\sim(\hbar/ND\kappa)$.

Going to the case of {\sl two open channels}, $a$ and $b$, we will have two
$N$-dimensional amplitude vectors, ${\bf B}^{a}$ and ${\bf B}^{b}$. In
accordance with that, in the regime of strong continuum coupling we find two
broad states, $|\Phi_{\pm}\rangle$, which divide among themselves the lion's
share of the total width $w$. Their complex energies, ${\cal E}_{\pm}=E_{\pm}-
(i/2)\Gamma_{\pm}$ depend on the amplitudes, $\gamma^{a,b}=2\pi|{\bf
B}^{a,b}|^{2}$, of the vectors ${\bf B}$ and the angle $\vartheta$ between them
\cite{SZAP92}. In particular, after the super-radiant transition, $\kappa\gg
1$,
\begin{equation}
\Gamma_{\pm}\approx\,\frac{1}{2}\,\left[w\pm\sqrt{w^{2}-4\gamma^{a}\gamma^{b}
\sin^{2}\vartheta}\right].                        \label{60}
\end{equation}
If the decay amplitudes are close to orthogonality, the channels become
effectively decoupled and we obtain two super-radiant states (the left and
right edge states in the lattice example). In the ``parallel" situation,
$\vartheta\rightarrow 0$, the state $|\Phi_{-}\rangle$ with destructive
interference of the two decay channels joins the set of trapped states, while
the constructive interference leaves only one super-radiant state
$|\Phi_{+}\rangle$ with $\Gamma_{+}\rightarrow w$. The long-lived $N-2$ states
are concentrated in the subspace orthogonal to that spanned by the vectors
${\bf B}^{a,b}$.

We can qualitatively anticipate the results for a {\sl multi-channel} case. The
matrix $W$ has a rank $M$ equal to a number of open channels. This
defines the number of its non-zero eigenvalues; for $M=1$, the only
nonzero eigenvalue of this matrix equals its trace $w$. Therefore, after the
transition to the strong continuum coupling, the natural basis would be that of
the {\sl doorway} representation, where $M$ eigenvectors of $W$ with
non-vanishing eigenvalues are used along with $N-M$ vectors from the
complementary subspace. Here we expect the appearance of $M$ broad states
sharing the total width, while $N-M$ states are getting trapped. Such a
picture was clearly observed in the numerical calculations
\cite{kleinwachter85} of nuclear structure of $^{16}$O in the region of the
dipole resonance studied as a function of the continuum coupling strength.
However, if the multidimensional angles $\vartheta^{ab}$ between the amplitude
vectors fluctuate, the possible destructive interference can suppress the
manifestations of super-radiance, especially in the situation of many open channels.

The doorway physics will be discussed in more detail later. Here we just
mention one simple but instructive model \cite{VZWNMP04}, a set of $n$
single-fermion levels $\epsilon_{j}$, including the upper one, $j=1$,
in the continuum with complex energy $\epsilon_{1}-(i/2)\gamma$. In the system
of $k<n$ fermions, the interaction mixes the configurations partially populating
the level $\epsilon_{1}$ and opening the decay for other states. If all
$n!/[k!(n-k)!]$ many-body states have energy higher than the decay threshold,
in the presence of the interaction they acquire the widths and shift down from
the real energy axis into the complex energy plane. However, with the increase of
the width $\gamma$, complex plane trajectories of some states turn back and again
approach the vicinity of the real axis while the rest of states have
monotonously growing widths, {\bf Fig. 1}. The number of the latter
(super-radiant) is $(n-1)!/[(k-1)!(n-k)!]$, the number of configurations of
remaining $(k-1)$ particles compatible with the occupied doorway state $|1)$
(these particles share the orbits $j\neq 1$).

The segregation of super-radiant and trapped states occurs in fact rather
sharply when the parameter $\kappa$ reaches the value close to 1. This is
illustrated by {\bf Fig. 2} taken from Ref. \cite{lehmann95} where the system of
1000 intrinsic states coupled uniformly to 250 channels was modeled. The
random distribution of the widths at weak coupling evolves releasing the
short-lived states until the cloud of super-radiant states sharply separates
from the rest of the system that contains trapped states with prolonged
lifetimes.

\newpage

\section{Nuclear continuum shell model}

As mentioned in the previous Section, the story of the effective non-Hermitian
Hamiltonian was essentially started with the search for an appropriate theory for
description of resonance nuclear reactions. Although we will argue that the
theory is much more universal and will show the examples of its broad
applications, still nuclear physics provides probably the most explored arena
for using this approach. At least partly, this can be explained not only by
historical reasons. The main trends of the development of modern nuclear
science clearly lead to the problems, where the continuum effects are
exceedingly important.

Experimental nuclear physics shifted its center of interest to the nuclei far
from the valley of stability. The new physical problems combine here the
questions of limits of nuclear existence with the astrophysical mysteries of
the origin of chemical evolution and supernova explosion. The standard shell
model is constructed to work in the discrete
spectrum of nuclear, atomic or molecular states. The beta-unstable nuclei close
to the drip lines are loosely bound, in many cases only the low-lying states
(or just the ground state) are particle-stable. Therefore the conventional
distinction between structure and reactions (the division between the subjects
of the research as well as between the researchers) becomes obsolete and it is
necessary to find ways to overcome this barrier. In fact, a similar
situation exists for the entire class of {\sl open} or {\sl marginally stable}
mesoscopic systems, including various micro- and nano-scale channels for
transmission of quantum information.

In loosely bound nuclei any real or virtual excitation is related to the
continuum part of the energy spectrum. The sharp discrimination between nuclear
structure and nuclear reaction disappears and our theoretical description
should cover both regions. In the classical problem of neutron resonances in
heavy nuclei we also are working in the continuum very close to the threshold
of the discrete spectrum. But there (we return later to this class of problems)
the details of the underlying structure are less important since the long-lived
resonances correspond to chaotic states of the compound nucleus, where one can
use statistical approaches justified by random matrix theory. In nuclei close
to the drip lines we are interested in details of weakly bound structure and in
the spectroscopy of individual resonances. Therefore we need the
generalization of the shell model that would allow the full understanding of
this region and unified description of structure and reactions.

Here the approach based on the effective non-Hermitian Hamiltonian becomes
extremely appropriate (along with other kindred methods, like for example the
Gamow shell model \cite{michel03,michel09}). The signatures of super-radiance
were seen (however not fully understood) in numerical simulations by Moldauer
\cite{moldauer68}. The first clear indication can be found in Ref.
\cite{kleinwachter85}, where the dipole excited states in $^{16}$O were
considered in a simplified shell model version, with decay probabilities (level
widths) taken as variable parameters. The results sharply changed with the
degree of openness of the system. Depending on the number $M$ of open
channels, as the escape widths were increased, the formation of $M$ very
broad resonances was observed, with other states becoming more and more
narrow. The analogy to optical super-radiance and the explanation of these
effects were suggested in Refs. \cite{SZPL88,SZNPA89}; the first results of
applications to the nuclear shell model were given in the review article by
Rotter \cite{rotter91}. Our discussion will be mainly based on the later
development \cite{rotter01,VZcont03,VZCSM05,CSM06}.

\subsection{Two-level example}

We start with the illustration of dynamics on the borderline between discrete
spectrum and continuum. We assume that the intrinsic, ${\cal Q}$, space
contains only two relevant states coupled to the single open decay channel
through real amplitudes $A_{1,2}$. This simple example \cite{VZcont03} in fact
has many non-trivial physical applications
\cite{rotter01,brentano96,brentano99,magunov99,magunov01,brentano02} including
experimental studies with microwave cavities \cite{philipp00}. It might serve
\cite{brentjolos02} as the starting model for such a nucleus as
$^{11}_{3}$Li$_{8}$, where the open channel contains the residual nucleus
$^{9}$Li and emitted neutron pair. For the pair we assume only one state
$^{1}S_{0}$ of relative motion. In some sense, this is an original Cooper effect
\cite{cooper56} where one pair can be bound on the background of other fermions
even if the primary attractive interaction within the pair is insufficient for
binding; the attraction in the vicinity of the nucleus is also helped by the
exchange of the collective excitations of the core \cite{brinkbroglia,potel09}.

The most energetically favorable wave function of the Cooper pair is a
combination of available for the pair orbitals, for example $0p_{1/2}$ and
$1s_{1/2}$ in our simplified model (in reality, probably, it has more
components). Both single-particle states are in the continuum since the
adjacent isotope $^{10}$Li is unbound. The effective Hamiltonian includes
interactions of both types, the usual Hermitian $v$, such as pairing, and
the anti-Hermitian, through the decay channel, and it can be written in
a general form (\ref{27}),
\begin{equation}
{\cal H}=\left(\begin{array}{cc}
       \epsilon_{1} -(i/2)A_{1}^{2} & v-(i/2)A_{1}A_{2} \\
v-(i/2)A_{1}A_{2}            & \epsilon_{2}-(i/2)A_{2}^{2}\end{array}\right),
                                                          \label{61}
\end{equation}
where we assume that the real energies $\epsilon_{1,2}$ are both positive
(above threshold that is set at zero energy), and the factor $\sqrt{2\pi}$ is
included into the amplitudes $A_{1,2}$.

The complex eigenvalues of this Hamiltonian are {\sl formally} given by
\[{\cal E}_{\pm}=\,\frac{\epsilon_{1}+\epsilon_{2}-(i/2)(\gamma_{1}+
\gamma_{2})}{2}\]
\begin{equation}
\pm \frac{1}{2}\sqrt{(\epsilon_{1}-\epsilon_{2})^{2}+4v^{2}-
\frac{(\gamma_{1}+\gamma_{2})^{2}}{4}-i[(\epsilon_{1}-\epsilon_{2})(\gamma_{1}-
\gamma_{2})+4vA_{1}A_{2}]},                              \label{62}
\end{equation}
where $\gamma_{1,2}=A_{1,2}^{2}$. This result is only formal if, as it happens
in the realistic shell model, the amplitudes $A_{1,2}$ are energy dependent and
have to vanish at threshold. A typical complication in such problems is that
the location of threshold is unknown {\sl a-priori}. A consistent solution
would require to consider simultaneously the daughter nucleus and maybe even
the chain of decays as we will show later.

The behavior of complex energies predicted by eq. (\ref{62}) is non-trivial
\cite{VZcont03,brentano96}. The {\sl Hermitian perturbation} $v$ leads to the
standard {\sl level repulsion}; in the absence of open channels, the lower root
$E_{-}$ is pushed down crossing the zero level at $v^{2}=\epsilon_{1}\epsilon_{2}$
and becoming bound. The intrinsic states are mixed and, if their widths were
very different, after the mixing they become more uniform, {\sl the widths are
attracted}. Contrary to that, the {\sl imaginary perturbation} in the second
order makes the levels to {\sl attract} each other, while the widths are
{\sl repelled}, and this is the beginning of the road to super-radiation.
As this process is determined by the ratio of widths to the level spacing, it
occurs immediately for degenerate levels, as in the ideal Dicke model. Indeed, for $\epsilon_{1}=\epsilon_{2}\equiv \epsilon$ and $v=0$, we obtain
\begin{equation}
E_{+}=\epsilon, \quad E_{-}=\epsilon-\,\frac{i}{2}\,(\gamma_{1}+\gamma_{2}),
                                                       \label{63}
\end{equation}
the super-radiant state absorbs the whole width while the remaining state is
stable although its energy is embedded into continuum. In the same way, for
any number of degenerate intrinsic levels, the real Hamiltonian
$\widetilde{H}$ is given by the unit matrix, and the energy splitting is fully
defined by $W$ that produces $M$ non-zero eigenvalues. Another new effect
is the possibility of crossing for the complex poles \cite{hernandez02},
analogous to the well known behavior of two coupled classical damped
oscillators \cite{nussenzveig72}.

The non-universal dependence of amplitudes $A_{1,2}$ [and even of real energies
$\epsilon_{1,2}$ if the principal value part $\Delta(E)$, eq. (\ref{29}), is
important] on running energy $E$ makes the analysis much more involved
\cite{SZAP92,VZcont03,CSM06}, since the complex poles move as a function of
energy, and the position $E$ of a resonance on the real energy axis is defined
by the solution of ${\tt Re}{\cal E}(E)=E$. The two levels in our problem are
then found by diagonalizing different matrices. One should carefully take care
of the vanishing width at threshold, so for example, when the lower level
becomes bound, $E_{-}=\Gamma_{-}=0$. In the model of $^{11}$Li, one can assume
the usual energy dependence from the potential model for $p$- and $s$-orbitals,
\begin{equation}
\gamma_{p}(E)=\beta E^{3/2}, \quad \gamma_{s}(E)=\alpha\sqrt{E}. \label{64}
\end{equation}
In  this way we obtain the self-consistent solution when, simultaneously with
$v^{2}=\epsilon_{p}\epsilon_{s}$, we come to $E_{-}=\Gamma_{-}=0$. For special
conspiracy of the parameters, the system also reveals a bound state embedded in
the continuum. In general, bound states, resonances and energy dependence of
the amplitudes are strongly correlated.

The same effective Hamiltonian determines the resonance part of the scattering
amplitude,
\begin{equation}
T(E)=\,\frac{E(\gamma_{1}+\gamma_{2})-\gamma_{1}\epsilon_{2}-\gamma_{2}\epsilon_{1}
+vA_{1}A_{2}}{(E-{\cal E}_{-})(E-{\cal E}_{+})},               \label{65}
\end{equation}
where we correct a misprint in the corresponding equation of \cite{VZcont03}.
Only in the degenerate case of eq. (\ref{63}), we see a clear Breit-Wigner
resonance on the super-radiant state,
\begin{equation}
T(E)=\,\frac{\gamma_{1}+\gamma_{2}}{E-\epsilon+(i/2)(\gamma_{1}+\gamma_{2})}.
                                                               \label{66}
\end{equation}
The dark state with $\Gamma=0$ is not seen in the scattering. {\bf Fig. 3}
taken from \cite{VZcont03} shows the typical cross sections for the positive,
3{\sl a}, and negative, 3{\sl b}, signs of the product $A_{1}A_{2}$ at certain
values of parameters. The peak at low energy in the case of the positive
product and $v^{2}> \epsilon_{1}\epsilon_{2}$, when the lower level is loosely
bound, demonstrates a typical example of the threshold maximum of sub-resonant
nature, whereas the remaining broad upper resonance at higher energy is not
shown. For the negative product $A_{1}A_{2}$, the presence of the threshold
singularity is responsible for the finite value of the cross section at zero
energy.

\subsection{Realistic calculations}

In all realistic applications, the truncation of both, ${\cal P}$ and ${\cal Q}$,
spaces is inevitable. With a truncated set of single-particle orbitals, the space
of many-body states in the nuclear shell model, built, with full antisymmetrization
required by Fermi-statistics, on these orbitals, can be naturally identified with
our ${\cal Q}$-space. The configuration interaction within this space is given by
a set of matrix elements, traditionally of two-body but in modern versions also
of three-body character \cite{ZYad09,volya09}. The solution process is usually based on
a large-scale diagonalization of the Hermitian Hamiltonian matrix. The single-particle
states can be conveniently defined in a spherically symmetric basis being
labeled by the spin-spatial quantum numbers ($n,\ell,j,m$) and the isospin
projection $\tau$. The diagonalization then leads to the many-body states
with good quantum numbers of total spin and parity. The deformed mean field
and corresponding rotational bands can emerge from the solution if the original
space is sufficiently large.

In this spherical shell model, the interaction matrix elements are defined for any
pair of initial and final two- or three-body states with the same constants of motion
$-$ total spin quantum numbers ($J,M$), parity, and isospin if there are no forces
breaking charge symmetry. The number of such allowed matrix elements can be large
but still by orders of magnitude smaller than the dimension of the ${\cal Q}$-space.
The values of the matrix elements can be in principle theoretically derived from
the vacuum nucleon and meson interactions fitted by few-body observables.
In practice, these values have to be phenomenologically adjusted in order to reproduce
thousands of experimental data. Here the theory has already a rich experience and
many successes. Unfortunately, such an experience is essentially absent for the matrix
elements in the $H_{{\cal PQ}}+H_{{\cal QP}}$ class for coupling of the intrinsic space
with the outside world.

In the first stage of the development, usually a potential of the Woods-Saxon type and
simple short-range residual forces were used in all parts of the effective Hamiltonian;
the open channels were restricted to the one particle in the continuum
\cite{barz77,rotter84}. The single-particle resonances emerge from the internal
shell-model bound states embedded in the open continuum. In a similar way,
but with a more advanced model of intrinsic dynamics, it was possible to describe
the inelastic electron scattering with the knock-out of a nucleon \cite{fladt88}.

Since the effective Hamiltonian is energy-dependent, in general
the position $E_{r}(E)$ and the width $\Gamma_{r}(E)$ of a resonance
are parameters of the eigenvalues (\ref{6}) of ${\cal H}(E)$ which depend on
running energy of the experiment. The strictly formulated theory is capable
of predicting directly the observable cross section $\sigma^{ab}(E)$ for given
entrance and exit channels. Then the convenient working definitions of
the resonance parameters are the roots of obvious equations, $E_{r}=E_{r}(E=E_{r})$
and $\Gamma_{r}=\Gamma_{r}(E=E_{r})$. The shape of the resonance curve is evolving
with the energy of the experiment being in general non-symmetric with respect to
the position $E_{r}$ \cite{barz77} and reveals a cusp required by unitarity at the
threshold for opening a new channel \cite{wigner48,baz57,breit57}. The cusps are well
known experimentally both in nuclear physics and particle physics and clearly seen
in exactly solvable models \cite{ahsan10} and in full modern continuum shell model
applications \cite{TDCSM,michel07}.

The routine identification of the resonance width with the product of a spectroscopic
factor (occupancy of a single-particle orbital) and penetrability is invalid being
approximately applicable only in the regime of separated resonances. In a more general
situation of strong coupling, the width results from repeated excursions to the continuum
and back. This should be especially important for many-particle processes as fission
and fusion of nuclei \cite{rotter91}. Partial widths for specific decay channels also
can be calculated, and in the regime of overlapping resonances, the total width $\Gamma_{r}$
is smaller than the sum of partial widths \cite{rotter84} while the equality holds for
well isolated resonances. Another important conclusion is that in general it is not correct
to arbitrarily parameterize the set of resonances as poles of the $S$-matrix. This can
violate subtle unitarity requirements which lead to correlations of positions and residues
at the poles, even for relatively well separated resonances \cite{mitchell86}.

In strongly bound nuclei, the spectroscopy of low-lying states can be decently
described without taking into account the continuum effects, although the
widths of resonances and the connection to the scattering problem are lost. In
nuclei far from stability the most interesting physics is just on the
borderline of the continuum and above. Here, the elements missing in the theoretical
foundation of the shell model are to be added phenomenologically, using general
physical arguments and available data. The situation is facilitated at low
energies and in light nuclei, where only few channels are open. Then a practically
reasonable approach is to add a phenomenological potential description to the
usual shell model inventory. In some versions of the shell model this is anyway
required if the starting point was a harmonic oscillator field and proper
radial characteristics are not defined by the model.

\subsubsection{Example: Oxygen isotopes}

A good example can be the shell-model description of the oxygen isotopes
\cite{VZcont03,VZCSM05,CSM06}. Of course there are promising developments of
technically different approaches to similar problems of shell model accounting for the
continuum effects \cite{rotter91,bennaceur99,okolowicz03}. Empirically, the last
particle-stable oxygen isotope is $^{24}$O. In the trivial shell model, this would
be a core $^{16}$O plus eight neutrons completely filling in the levels $0d_{5/2}$
and $1s_{1/2}$. The spin-orbit partner $0d_{3/2}$ is already in the continuum
(the neutron width of the $3/2^{+}$ state in $^{17}$O is 96 keV), while the intermediate
level $1s_{1/2}$ is weakly bound, if the level scheme is built on the base of $^{16}$O.
At low energy, and due to the extra stability of the doubly magic $^{16}$O, it is
reasonable to limit our ${\cal Q}$-space by this $sd$-shell model. The relevant
decay channels at low energy are single-neutron and two-neutron decays moving
the system down along the same oxygen isotope chain. As we mentioned, the
knowledge of the decay thresholds is rather critical. To know them we have to
study the spectroscopy of the daughter nucleus, and hence to look at the whole
chain in its entirety, all neutron-rich oxygen isotopes from the root $^{16}$O.
The two-neutron channels can be restricted by the emission of the correlated pair
in the $^{1}S_{0}$ state. It is important to have in mind that for stable states,
when the decay widths vanish, this approach automatically reduces to the routine
shell model without continuum, maybe with renormalized parameters due to the shift
operator $\Delta(E)$.

In practice we need to add to the standard $sd$-model formulation, see for example
\cite{USD,HBUSD} new parameters describing the two neutron widths under an assumption
that only the lowest states of allowed seniority (equal to zero or one) are populated
in the daughter nucleus. For excitation energy above $\sim 4$ MeV, one needs to include
the final states with higher seniority. The widths for the emission of a $d$-neutron
are supposed to grow from the corresponding threshold being proportional to the energy
excess to the power 5/2 characteristic for $\ell=2$. The coefficients are found from
the neutron scattering problem off the $^{16}$O nucleus. Even in the over-simplified
framework where the residual interactions were limited to their main components,
pairing and monopole terms \cite{VZcont03}, the results were quite satisfactory
indicating the rich perspectives of the whole approach. We have to mention that
in principle the effective shell-model interaction could be also complex in the presence
of open decay channels; this problem was not solved yet. In calculations of reaction
cross sections, the potential phases, especially Coulomb for charged projectiles, can be
included along with the dynamics based on the coupling Hamiltonian $H_{{\cal PQ}}+H_{{\cal QP}}$.

The rediagonalization of the effective Hamiltonian for daughter nuclei accounts for
the structural rearrangement of the remaining nucleus after neutron decay. {\bf Fig. 4}
\cite{VZCSM05} shows the results of the continuum shell model for the chain of oxygen
isotopes on the base of the shell model interaction \cite{HBUSD}. There is a good qualitative
and often quantitative agreement with available experimental data. Among interesting
predictions, one can mention the ground state width of only 112 keV for the first
particle-unbound even oxygen isotope, $^{26}$O, where the $Q$-value for the two-neutron
decay to the last bound isotope $^{24}$O is predicted to be only 21 keV. In this case only
sequential two-neutron decays were considered; this should be extended for correlated
pair decays.

\subsubsection{Some technicalities}

A favorable technical feature of working with the effective Hamiltonian is the absence of
the need to use explicitly the complex scaling and related tools. One can still keep
a diverse experience of the routine shell model and introduce new computational methods
for the non-Hermitian Hamiltonian problem \cite{TDCSM}.

One of such technical improvements is the propagator calculation using the time representation
of the propagator (\ref{33}) for ${\cal E}\rightarrow E^{(+)}$ as a Fourier image for positive
time of the Hermitian evolution operator that, in turn, is expressed as an expansion over
{\sl Chebyshev polynomials} \cite{ikegami02}, $T_{n}(\cos x)=\cos(nx)$,
\begin{equation}
e^{-iHt}=\sum_{n=0}^{\infty}(-i)^{n}(2-\delta_{n0})J_{n}(t)T_{n}(H).         \label{67}
\end{equation}
The main advantage of the representation is in the long-time dynamics where the Bessel
functions $J_{n}(t)$ lead to the convergence important for the energy resolution in the following calculation of the cross sections, strength functions and level densities.
Such an approach avoids direct diagonalization of large complex matrices.

With the constructed Green function (\ref{33}) at real energy and for energy-independent
shell-model part $H_{{\cal QQ}}$ of the Hamiltonian, one can solve for the full propagator
(\ref{32}) using the Dyson equation or its equivalent (\ref{34}) (the so-called {\sl Woodbury
equation}) where the coupling with and through the continuum includes only the matrix in
the channel space, usually of a much smaller dimension compared to the intrinsic shell-model space. The full Green function ${\cal G}$ allows one to study the damping of quasistationary intrinsic states with non-exponential decay of broad resonances. This approach, supplemented
with addition of the collectivizing Hermitian interaction that is absent in the standard $sd$-shell model, also gives a possibility to study giant resonances in loosely bound nuclei,
a subject of great interest and contradictory predictions. In this approach the principal value term of self-energy (\ref{29}) can be included without difficulty and, in practical applications \cite{TDCSM}, it improves the description of resonances seen \cite{hoffman09} in neutron scattering from exotic oxygen isotopes.

\newpage

\section{Doorway states}

\subsection{Dynamics based on doorway states}

In a complicated system, it is often the case that only a {\sl subset} of intrinsic
states $|q\rangle$ of space $\{{\cal Q}\}$ connects directly to the $\{{\cal P}\}$ space
of channels. The rest of states in $\{{\cal Q}\}$ will connect to $\{{\cal P}\}$ only
when they acquire admixtures of these selected states of the first type. The special
states directly coupled to continuum are the {\sl doorways}, $|d\rangle$, see for example
\cite{feshbach67}. They form the doorway subspace $\{{\cal D}\}$ within $\{{\cal Q}\}$,
and the corresponding projection operator will be denoted here as ${\cal D}$.
The remaining states in $\{{\cal Q}\}$ will be denoted as $|\widetilde{q}\rangle$
and their subspace as $\{\widetilde{{\cal Q}}\}$.

The full Hamiltonian in Eq. (\ref{19}) can be decomposed now as
\[H=\left(H_{\widetilde{{\cal Q}}\widetilde{{\cal Q}}}+H_{{\cal D}{\cal D}}+
H_{\widetilde{{\cal Q}}{\cal D}}+H_{{\cal D}\widetilde{{\cal Q}}}\right)\]
\begin{equation}
+\left(H_{{\cal P}{\cal P}}+H_{{\cal D}{\cal P}}+H_{{\cal P}{\cal D}}\right).
                                                              \label{68}
\end{equation}
Note that the terms $H_{{\cal P}\widetilde{{\cal Q}}}$ and $H_{\widetilde{{\cal Q}}
{\cal P}}$ are missing because in accordance with the doorway hypothesis they are
negligible. Also note that the diagonalization of the Hamiltonian part in the upper
line of eq. (\ref{68}) would provide the states $|q\rangle$ with the components of
$|d\rangle$ mixed with the $|\widetilde{q}\rangle$ states. The two last items in
the above equation couple the doorway states, and therefore also the $|q\rangle$
states to the open channels.

Doorway states play an important role emphasizing collective effects in description
of scattering and reactions with many-body systems. They are often serving as
the connection between the continuum and the dense spectrum of quasi-bound states.
As we will now see, the concept of the doorway is, in certain situations, closely
connected to the notion of super-radiance \cite{auerbach07}.

\subsubsection{The case of a single doorway}

Let us consider the case when there is only one important doorway state $|d\rangle$.
The matrix elements of the effective operator $W$, eq. (\ref{30}), in the intrinsic
space are now given by
\begin{equation}
\langle q|W|q'\rangle=2\pi \sum_{c({\rm open})}\langle q|H_{{\cal DP}}|c\rangle
\langle c|H_{{\cal PD}}|q'\rangle                        \label{69}
\end{equation}
with the doorway assumption:
\begin{equation}
\langle q|H_{{\cal DP}}|c\rangle=\langle q|d\rangle\langle d|H_{{\cal DP}}|c\rangle,
                                                         \label{70}
\end{equation}
where $\langle q|d\rangle$ is the amplitude of the admixture of the doorway into
the $|q\rangle$ state.  Eq. (\ref{70}) now becomes:
\begin{equation}
\langle q|W|q'\rangle=2\pi\langle q|d\rangle\langle d|q'\rangle \sum_{c({\rm open})}
\Bigl|\langle d|H_{{\cal DP}}|c\rangle\Bigr|^{2},          \label{71}
\end{equation}
where again we have separable matrix elements of the matrix $W$, this time irrespective
of the number, $M$, of open channels. The doorway serves as a {\sl filter} for
the coupling of the $\{{\cal Q}\}$-space to the space of open channels $\{{\cal P}\}$.
As a result, we find one broad state with a width
\begin{equation}
\Gamma_{d}^{\uparrow}=2\pi\sum_{q}|\langle q|d\rangle|^{2}\,\sum_{c}
\Bigl|\langle d|H_{{\cal DP}}|c\rangle\Bigr|^{2} =
2\pi\sum_c\Bigl|\langle d|H_{{\cal DP}}|c\rangle\Bigr|^2.   \label{72}
\end{equation}
This width is nothing but the {\sl decay width of the doorway} into all available channels.

\subsubsection{When is this approach valid?}

As discussed above, the criterion of validity of the super-radiant mechanism is that
the average spacing between the levels in $\{{\cal Q}\}$-space is smaller than the
decay width of a typical state ``before" the super-radiant mechanism takes effect.
This can be expressed, in the case of doorways, in the following way \cite{auerbach07}.

Consider the {\sl spreading width}, $\Gamma_{d}^{\downarrow}$, of the doorway state
representing the fragmentation of $|d\rangle$ into compound states $|\widetilde{q}\rangle$.
If $N_{q}$ is the number of compound states in the interval covered by the spreading
width, their average energy spacing is
\begin{equation}
D_{q}\approx \,\frac{\Gamma_{d}^{\downarrow}}{N_{q}}.      \label{73}
\end{equation}
Before the super-radiant mechanism is turned on, the average decay width of a typical
$|q\rangle$ state is
\begin{equation}
\Gamma_{q}^{\uparrow}=2\pi \sum_{c}\Bigl|\langle q|H_{{\cal QP}}|c\rangle\Bigr|^{2},
                                                                \label{74}
\end{equation}
which can be estimated by
\begin{equation}
\Gamma_{q}^{\uparrow}\approx \,\frac{\Gamma_{d}^{\uparrow}}{N_{q}}.      \label{75}
\end{equation}
These estimates result in
\begin{equation}
\frac{\Gamma_{q}^{\uparrow}}{D_{q}}\,\approx\,\frac{\Gamma_{d}^{\uparrow}}
{\Gamma_{d}^{\downarrow}}.                                   \label{76}
\end{equation}

We conclude that the requirement for the super-radiant doorway mechanism to be valid
can be formulated as
\begin{equation}
\frac{\Gamma_{d}^{\uparrow}}{\Gamma_{d}^{\downarrow}} >1,      \label{77}
\end{equation}
the decay width of the doorway should be comparable or larger than its spreading width
in order for the super-radiant mechanism to be valid. This condition is not always
obeyed, but in some cases, as we will see, it is satisfied.

\subsubsection{Isobaric analog resonances}

An outstanding example, where the condition (\ref{77}) is fulfilled and the super-radiant
mechanism may take place, is given by the isobaric analog state (IAS) $|IAS\rangle$
\cite{AUE72} that is obtained from the parent state $|\pi\rangle$ with certain isospin $T$
by acting with the isospin lowering operator $T_{-}$ that changes a neutron into a proton:
\begin{equation}
|IAS\rangle=\,{\rm const}\cdot T_{-}|\pi\rangle.             \label{78}
\end{equation}

In medium and heavy nuclei, the IAS are surrounded by many compound states $|q\rangle$
of lower isospin $T_{<}=T-1$. The Coulomb interaction violates the isospin symmetry
fragmenting the strength of the IAS over many states $|q\rangle$ and giving rise to
the spreading width $\Gamma_{IAS}^{\downarrow}$ of the IAS. If located above thresholds,
the IAS can also decay into several continuum channels with the decay width
$\Gamma_{IAS}^{\uparrow}$. In such cases the condition (\ref{77}) is usually satisfied.
For  example, in $^{208}$Pb the spreading width of the IAS is about 80 keV while
the escape width is about 160 keV \cite{AUE72}, thus the ratio
$\Gamma_{IAS}^{\uparrow}/\Gamma_{IAS}^{\downarrow}\approx 2$. This situation is typical
for other heavy and medium mass nuclei. In the $Z\sim 40$ region the above ratio is
in some cases even larger than 2 \cite{AUE72}. The super-radiant mechanism is therefore
relevant to this case providing a straightforward explanation why the IAS appears as
a single resonance with the decay width of the doorway $|IAS\rangle$,
\begin{equation}
\Gamma_{IAS}^{\uparrow}=2\pi\,\Bigl|\langle IAS|H_{{\cal QP}}|{\cal P}\rangle\Bigr|^{2}.
                                                           \label{79}
\end{equation}

\subsection{Giant resonances}

Giant resonances in nuclei (or atomic clusters) present another example of similar
physics \cite{AK83, BBBGR}. Usually, the giant resonances are discussed in terms of
particle-hole ({\sl p-h}) configurations with identical spin-parity quantum numbers.
The coherent residual interactions form a correlated state that carries much of
the transition strength of a corresponding multipole operator. The giant resonances
are mostly located in the particle continuum decaying via particle emission to
the ground and excited states in the daughter nucleus.

Since the $p-h$ giant resonance $|G\rangle$ is surrounded usually by a dense spectrum
of $2p-2h$ and more complex configurations, the residual strong interaction will mix
the state $|G\rangle$ with this background. Each of the resulting states denoted as
$|b\rangle$ will contain the admixture $\langle G|b\rangle$ of the giant resonance.
When above the threshold, the mixed states $|b\rangle$ couple to the continuum.
If we assume that the dominant coupling is via the admixture of the giant state,
then $|G\rangle$ serves as a doorway, and the matrix $W$ is separable, given by
\begin{equation}
\langle b|W|b'\rangle=\langle b|G\rangle\langle G|b'\rangle\,\sum_{c}
\langle G|A|c\rangle\langle c|A|G\rangle,                       \label{80}
\end{equation}
where $\langle G|A|c\rangle$ are the decay amplitudes for the giant resonance state.
The matrix $\langle b|W|b'\rangle$ is again of rank one and the non-zero eigenvalue
is equal to
\begin{equation}
\Gamma_{G}^{\uparrow}=2\pi\,\sum_{c}\Bigl|\langle G|A|c\rangle\Bigr|^{2}.
                                                   \label{81}
\end{equation}
This is valid under the assumption that the energy spread of the background states
$|b\rangle$ is small compared to their decay width,
\begin{equation}
\Gamma_{b}^{\uparrow}=2\pi\,\sum_{c}\Bigl|\langle b|A|c\rangle\Bigr|^{2}.
                                                \label{82}
\end{equation}
As before, this condition can be expressed as
\begin{equation}
\frac{\Gamma_{G}^{\uparrow}}{\Gamma_{G}^{\downarrow}}\,>1,   \label{83}
\end{equation}
the spreading width of the giant resonance is smaller than its total decay width.

When the condition (\ref{83}) is not satisfied, so that the energy intervals
between the background states are larger than their decay widths $\Gamma_{b}^{\uparrow}$,
the situation of a single decay peak for the giant resonance might not hold.
Still one could expect some bunching of background states into groups with spacing
within the group smaller than the decay widths. Each such a group then can be treated
separately using the super-radiant mechanism and appear as a single peak in the decay
(or excitation) curve. One will then observe {\sl intermediate structure} resonances
in the giant-resonance energy domain.

One should remark that for many giant resonances, such as the giant dipole resonance
(GDR) or giant isoscalar quadrupole resonance, the above condition (\ref{83}) is not
fulfilled \cite{AK83,BBBGR}, and the spreading width of the resonance exceeds the decay
width. However, in continuum RPA calculations it was found that high lying giant
resonances have substantial decay widths \cite{AK83}, larger than their spreading widths.
For example, the isovector monopole resonances calculated in heavy nuclei had decay
widths of the order of 8-10 MeV, larger than the spreading widths which are of the order
of a few MeV. In these cases the condition (\ref{82}) for super-radiance is satisfied.

The GDR in nuclei not far from the valley of stability is shifted up in energy compared
to the unperturbed excitation energy $\sim 1\hbar\omega$ of $p-h$ states. The GDR strength
then saturates a significant part of the classical dipole sum rule although some strength
still remains around the unperturbed energy. The situation changes for nuclei with
a large neutron excess farther away from stability. The fraction of dipole strength in
$p-h$ states that are not shifted increases due to the possibility of cooperative
oscillation of the neutron excess against the core. If in an unstable nucleus some of
these states are already in the continuum, they interact not only by the residual
dipole-dipole interaction but through common decay channels. Then the situation reminds
what we have discussed for the main branch of the GDR. As a result, the so-called
{\sl pygmy-branch} of the giant resonance \cite{adrich05,tonchev10} is formed being currently a subject of serious experimental interest. This soft collective dipole mode serves again as a doorway state. Such a mechanism of formation of the pygmy-resonance was predicted in \cite{SZZagreb} and considered more in detail in \cite{ZVVarenna}.

In the limit of small energy spread of the $p-h$ states, the interplay and redistribution
of dipole strength and decay width to a channel $c$ between the main GDR and the pygmy-resonance
is determined by the ``angle" between the multidimensional vectors ${\bf A} =\{A_{q}^{c}\}$
of decay amplitudes and ${\bf d}=\{d_{q0}\}$ of dipole matrix elements connecting $p-h$
excitations and the ground state. If these vectors are ``parallel", the soft mode loses its
collective character, and the main GR peak concentrates both the strength and the continuum width.
A situation of this type takes place in the case of the collective $\Delta$-isobar$-$nucleon-hole
states, discussed in the Section 5.2, when the pion matrix elements determine both the strength
and the width \cite{AZ02,hirata79} of the doorway state. In the extreme ``orthogonal" case
the main peak would concentrate the multipole strength but the whole width would go to the soft
mode so that the main peak would become a ``zombie" state embedded in the continuum.
The reality corresponds to an intermediate situation with the weight of the pygmy-branch rising
closer to the drip line. In such cases it makes sense to speak about {\sl exit doorways}
\cite{adhikari83,canto94}.

\subsection{Several channels and intermediate structure}

Let us turn to the situation when the $|q\rangle$ states are coupled to a number of open
channels in the ${\cal P}$-space, a typical situation in nuclear physics. In addition to
the elastic channel, the compound states in many cases can decay to excited states in the daughter
nuclei. In this situation we have a number of open channels $M$ that is still much smaller
than the number of $|q\rangle$ states. In many cases it is not the same doorway that connects
to all the open channels. In fact often the doorways are specified, and their number is
equal to the number of open channels \cite{auerbach07}.

We should point out the difference between the situation mentioned here and the case of
a single doorway for all the channels discussed above. In the latter case the matrix $W$
was factorized, eq. (\ref{70}), as $W_{qq'}=a_{q}a_{q'}^{\ast}$ but the present case is
more complicated because each channel has its own doorway, so that the matrix $W$ is of
the form
\begin{equation}
W_{qq'}=\,\sum_{d=1}^{M}a_{q}^{d}a_{q'}^{d\ast}.              \label{84}
\end{equation}
When the number $|q\rangle$ states is much larger than the number of channels $M$,
correlations are present in the matrix $W_{qq'}$. In this case the cross section will
exhibit several resonances, with intermediate widths $\Gamma_{{\rm int}}$ such that
$\Gamma_{q}\ll\Gamma_{{\rm int}}<\Gamma_{{\rm s.p.}}$, where $\Gamma_{{\rm s.p.}}$ is
the decay width of a single-particle resonance.  Such intermediate resonances were
observed in nuclear reactions long ago \cite{feshbachdoorway,auerbach07} and the cross
sections were said to exhibit intermediate structure.

In some situations, several doorway states $|d\rangle$ appear separated by energy intervals
larger than their total widths. The intrinsic states $|q\rangle$ couple to these doorways
locally giving rise to spreading of doorways. According to the doorway hypothesis, the
$|q\rangle$ states couple to the continuum only via the admixture(s) of the local doorways.
Then the matrix $W$ separates into several disconnected quadratic blocks of dimensions
$n_{d}\times n_{d}$, where $n_{d}$ is the number of states $|q\rangle$ contained within
the spreading width of the doorway $|d\rangle$. For each such a sub-matrix, eq. (\ref{71})
is valid, and the problem is reduced to that of the single doorway. The resulting picture
is a series of resonances separated in energy with decay widths equal to the decay widths of
the individual doorways.

\subsection{Highly excited doorway state}

As excitation energy grows, many open decay channels are getting accessible. This region
has its own specific features. A doorway state is singled out by a specific ``signal", either
a quantum number different from that of the majority of background states, or by a character
of intrinsic motion and, correspondingly, a type of external excitation.

A nuclear example of the first kind is given by the IAS that is a state of isospin $T_{>}
=T$ that is embedded into a sea of states of isospin $T_{<}=T-1$. The Coulomb interaction and
a small component of nuclear forces violating charge independence mix the IAS to the
background. This {\sl spreading width} of the IAS increases with excitation energy and then
saturates \cite{lauritzen}. If at this excitation energy the lifetime of the $T_{>}$ state
with respect to the decay outside is shorter than the lifetime of the background $T_{<}$ states,
the IAS preserves its individuality \cite{brentano82,melzer85}. One could expect that the purity
of the IAS will rapidly be washed out at high density of the admixed states. As it turns out,
just opposite is correct: after increased mixing, at some energy the purity of the IAS is
{\sl restored} as confirmed experimentally \cite{snover93}. In fact, this phenomenon was
predicted long ago by Morinaga \cite{morinaga55} and Wilkinson \cite{wilkinson56}.
A qualitative explanation follows from plausible kinetic arguments \cite{4B91}. The fully
quantum-mechanical doorway theory \cite{SZsimple} naturally justifies this conclusion.

Let the doorway state $|d\rangle$ be immersed into a rich set of $N\gg 1$ background states
$|\nu\rangle$. The state $|d\rangle$ with complex energy ${\cal E}_{0}=\epsilon_{0}-(i/2)
\gamma_{0}$ can decay into continuum (the corresponding width is $\gamma_{0}$) and to be
mixed with background states $|\nu\rangle$ by the coupling matrix elements $V_{\nu}$
which can be considered as real. In turn, the background states, after diagonalizing their intrinsic coupling, can at this energy decay into continuum through many, $M\gg 1$, complicated and essentially uncorrelated {\sl evaporation} channels; their complex energies are $\epsilon_{\nu}$. Because of the complexity of those states and presence of many open channels,
the coherent super-radiant collectivization of the background states is suppressed.
The coupling of the doorway state with the background defines new quasistationary states $|j\rangle$, and, similarly to the usual theory of bright stationary states, the complex
energies ${\cal E}_{j}$ are the roots of
\begin{equation}
{\cal E}_{j}-{\cal E}_{0}-\sum_{\nu}\,\frac{V_{\nu}^{2}}{{\cal E}_{j}-\epsilon_{\nu}}=0.
                                                                \label{85}
\end{equation}
Each root $j$ with energy ${\cal E}_{j}=E_{j}-(i/2)\Gamma_{j}$ carries a fraction $f_{j}$ of
the doorway strength, $\sum_{j}f_{j}=1$. Substituting the evaporation widths by their typical
value $\gamma_{{\rm ev}}$ and calculating the imaginary part of eq. (\ref{85}), we find
\cite{SZsimple}
\begin{equation}
f_{j}=\,\frac{\Gamma_{j}-\gamma_{{\rm ev}}}{\gamma_{0}-\gamma_{{\rm ev}}}. \label{86}
\end{equation}
This result has a clear probabilistic, or kinetic, meaning indicating two routes of each fine
structure resonance to the irreversible decay,
\begin{equation}
\Gamma_{j}=\gamma_{0}f_{j}+\gamma_{{\rm ev}}(1-f_{j}).                 \label{87}
\end{equation}

The further analysis shows the resulting picture and its evolution with excitation
energy from the viewpoint of reaction cross sections and time delays. The characteristic parameter here is the ratio $\gamma_{{\rm ev}}/D\sim N\gamma_{{\rm ev}}/\Gamma^{\downarrow}$
of the evaporation width to the spacing of compound states; the spreading width of
the doorway state is $\Gamma^{\downarrow} \sim ND$. Instead of super-radiance in this
situation of many uncorrelated open channels, when the background states overlap, the fluctuations in the excitation of the doorway state are washed away, see {\bf Fig.} 5,
and the depletion of the admixed background states of opposite isospin occurs faster
than their population through the doorway mechanism. This justifies the predictions
of isospin purification \cite{morinaga55,wilkinson56} observed in \cite{snover93}.
Such effects can be also interpreted as related to the quantum Zeno effect \cite{facchi02}.
In kinetic language of Ref. \cite{4B91}, this stage corresponds to temperature of
the compound nucleus exceeding a critical value when the total evaporation width of
the admixed states becomes of the order of the spreading width $\Gamma^{\downarrow}$
of the IAS. In a similar formalism, analogous hierarchical effects were considered
\cite{silvestrov01,amir08} for chaotic quantum dots or atoms coupled to the electromagnetic modes of a cavity immersed in a bigger box. At random coupling, one can see mesoscopic
fluctuations and oscillations of the survival probability.

Another related effect is the so-called {\sl disappearance of collective strength} of
giant resonances excited in a high-energy heavy ion collision \cite{gaard92}. The analogs
of giant resonances are expected to exist being built on all other nuclear states ({\sl Axel-Brink hypothesis}). However, beyond a critical excitation level, excited in heavy
ion collisions fine structure states, according to the mechanism discussed above, decay
in the continuum through incoherent evaporation channels {\sl before} they are capable to transfer excitation to a collective degree of freedom: in a sense, the original excitation
has no time to open the door through the doorway.

\subsection{Additional comments}

We have considered few examples justifying a generic qualitative picture of formation of
doorway states and intermediate structures in various nuclear processes. It would be interesting
to extend this approach to other mesoscopic systems (the application of the doorway concept
to giant resonances in atomic clusters was made in \cite{hussein00}). The notion of the
doorway state has been applied to atoms and molecules \cite{kawata00}, to quantum dots
\cite{baksmaty08} and fullerenes \cite{laarman07,hertel09} and recently even to a classical
phenomenon of earthquakes \cite{franco11}. In the physics of cold atoms \cite{SPBEC},
the coupling of the internal states to the continuum via the Feshbach resonance influences
profoundly the system affecting not only the imaginary part of the effective Hamiltonian but
also the real part. This enables one to manipulate the properties of the system in a trap by
changing the sign and/or the strength of the two-body interaction. The Feshbach resonance in
this case serves as a doorway.

The traditional theory of intermediate structure in nuclear reactions \cite{griffin66,FKK80}
considers usually statistical sequences in time of multi-step particle interactions with possible
pre-compound (or pre-equilibrium) decay to the continuum. The present formulation emphasizes
complementary aspects of underlying physics, namely those that follow from the strict
quantum-mechanical description of complicated many-body dynamics generated by the mean field
and residual interactions. The approach based on the effective non-Hermitian Hamiltonian
rigorously predicts the existence of different scales in the energy dependence of observed
cross sections. These scales are defined by the intrinsic dynamics that combines collective
(coherent) and chaotic (incoherent) features. The new collectivity responsible for the emergence
of the doorways and related intermediate structure appears, under certain physical conditions,
due to the factorized nature of corresponding terms in the effective Hamiltonian. This separability
is deeply related to the properties of unitarity of the scattering matrix. In this way we obtain
a supplementary description of many-body dynamics.

\newpage

\section{Other applications}

\subsection{Atomic and molecular physics}

Many-body physics of complex atoms has many features in common with nuclear
many-body physics $-$ shell structure, residual interactions, growth of
complexity of excited states with excitation energy, very complicated
$-$ ``compound" $-$ wave functions of mixed states with several excited
electrons, presence of the continuum etc. This structure and high level
density are revealed, in particular, in the photoabsorption spectrum,
see for example, \cite{griesmann88,connerade90}. Typical autoionizing
states appear already in the set of two-electron excitations.

%\subsubsection{Autoionizing states}
The whole super-radiant formalism was translated to the language of atomic physics in
Ref. \cite{FGG96}. The physical difference from analogous nuclear phenomena is mainly
due to the long-range character of the electrostatic interactions. In nuclear neutron
resonances typical spacings between the resonances are greater than their widths, and
the super-radiant collectivization requires higher energy above the neutron separation
threshold, where the effect might be overshadowed by the opening of new incoherent
channels. In atoms the influence of threshold factors, $\sim kR$, is suppressed by
the Coulomb effects, which facilitates the appearance of collectivization through
continuum. According to \cite{FGG96}, the photoabsorption experiments will be mostly
sensitive to narrow trapped states; in atoms the intrinsic collectivity similar to that
generating giant resonances in nuclei is substantially weaker. But in electron scattering
from ions one should observe broad doorway states and, with sufficient resolution, narrow
trapped states as well. This difference is clearly seen in numerical calculations
\cite{FGG96}.

Experimental data mainly received with multiphoton laser excitation of atomic beams
confirm the trend to the formation of trapped states with the autoionization rate
suppressed compared to the predictions for a simple atomic configuration
\cite{leeuwen94,luc95}. Ref. \cite{aymar96} provides a detailed review of this
research area.

The practical calculations of atomic spectra in principle can be done with the aid
of the same shell-model machinery (currently called by the term ``configuration
interaction" coined in quantum chemistry) as in nuclei, especially because the
interaction is essentially known. A new element comes from the presence in the same
energy region, along with complicated compound states, also the Rydberg states of
simple nature. The Rydberg spectrum is condensed near every ionization threshold.
However, the large difference in structure makes the mixing between compound
autoionizing states and Rydberg autoionizing states weak being in addition strongly
suppressed by their very different radial behavior. The real spectrum consists
therefore of two weakly mixing subsets.

Interesting physics appears in the case of a simple (for example, hydrogen) atom in
a strong uniform magnetic field \cite{friedrich85}. The bound states in the continuum
can be found according to the results of our two-level problem, Sec. 4.1. In the field
around the critical value, $B_{{\rm cr}}=m^{2}e^{3}c^{3}\hbar^{3}\sim 10^{5}$T,
the cyclotron energy $\hbar\omega_{{\rm c}}$, and therefore, the splitting of
Landau levels, are of the same order as atomic Rydberg energy $me^{4}/\hbar^{2}$.
The Rydberg series overlap, and the sequences of Landau levels in the continuum
interfere. This creates the situation of super-radiation and trapping, and, for
some values of parameters (magnetic field), bound states in the continuum.
It was stressed \cite{friedrich85a} that this phenomenon, earlier found in specifically
adjusted potentials, is more general leading to segregation of fast decaying and
trapped states.

In atomic physics one can directly calculate the necessary matrix elements between
the internal wave functions and continuum decay channels. An approximate statistical
treatment \cite{FGG96} of the states $J^{\pi}=4^{-}$ in the cerium atom showed that
these matrix elements have the Gaussian-type distribution that leads to the Porter-Thomas
distribution (PTD) of the resonance widths. Later, Section 7, we discuss the recent
experimental indications and corresponding theoretical arguments for nuclear neutron
resonances which show generic deviations from the PTD. In atoms the presence of two
components of the spectrum may distort the picture.

%\subsection{Molecular physics}
Radiationless and multiphoton transitions in atoms and molecules present an appropriate
applications of super-radiant physics. The effect of super-radiance and trapping
in molecular transitions in the approximation of degenerate intrinsic levels
coupled to the same decay channel was considered in \cite{verevkin88} and possible
generalization to non-degenerate levels was mentioned (called in this paper
``quasitrapping"). The actual presence of fast and slow dynamics in
the multiphoton excitation was noticed long ago \cite{beers75}.

\subsection{Intermediate and high energy physics}

As discussed in previous sections the super-radiant mechanism was applied to
a number of different systems involving many-body physics. The phenomena discussed
in the field of nuclear physics involved low energies. The universality of
the super-radiant mechanism leads us to consider also the field of intermediate
energy phenomena. The first discussion concerning this can be found in Ref.
\cite{auerbach94}. In this work the super-radiant mechanism was applied to
a number of situations involving intermediate energy hadronic systems.
The main objective of these studies was to point out that narrow resonances
can be formed due to the super-radiant mechanism. Here we bring some examples.

\subsubsection{The $\bar{N}N$ states}

In current relativistic theories \cite{horowitz81} of nuclear structure
the Dirac vacuum of occupied negative energy states plays an important role
in describing some of the nuclear properties. The influence of these negative
energy states, $\bar{N}$, is often implicit. It was suggested \cite{auerbach86}
that such theories could be tested by an observation of particle-hole states,
in which a nucleon is raised from the negative energy Dirac sea and placed in
an unoccupied state with positive energy. That we call an $\bar{N}N$ excitation.
In most relativistic models the nucleon occupying a negative energy state
experiences a strong potential of the order of several hundred, up to 700-800, MeV.
Thus the excitation energy of the $\bar{N}N$ state is much lower than the mass
of two nucleons, $E_{\bar{N}N}<2M_N$. The observation of such sub-threshold
nucleon-antinucleon excitations would provide a direct test of the relativistic
theories of nuclei. A question should be raised about the feasibility to observe
such excitations because of the expected large annihilation width. The $\bar{N}N$
pair is annihilated into mesons, in particular into several pions. One can expect
that the $\bar{N}N$ states will couple strongly to the annihilation channel and
decay very quickly. The large widths of these states will hamper their observation.

It is worth noting however that conditions for the formation of a super-radiant
state exist in this case \cite{auerbach86}. The spectrum of negative energy states
is within a given shell. (The spin-orbit splitting is nearly zero for the negative
energy states). The unoccupied positive energy states form also a dense spectrum,
and therefore one expects a strong degeneracy among the unperturbed $\bar{N}N$
configurations. The energy splitting between various $\bar{N}N$ states is typically
5-10 MeV, thus smaller than the individual decay widths into the common annihilation
channel. We are therefore in the regime of overlapping resonances.

Using a relativistic random phase approximation (RPA), a calculation of energies
and widths of $\bar{N}N$ states was performed in the framework of the $\sigma-\omega$
model \cite{huillier90} for $^{16}$O. The width was obtained from an imaginary
term in the meson propagators that mimicked the coupling to the annihilation
channel. The result of the diagonalization of the complex RPA matrices is quite
amazing. All the states that resulted from the diagonalization had width of the order
of 1-2 MeV, while one state had a width of about 160 MeV. This is the super-radiant
state. What happens here is that each $\bar{N}N$ state decaying to the common channel
produces width of the order of 15 MeV. As a result of the diagonalization of the matrix,
the wide super-radiant state emerges while the rest of states become very narrow and
thus live a long time. One could anticipate the following situation. In a simple
reaction, as for example nucleon pick-up, the super-radiant resonance, because of
its large width will show up as a flat bump contributing to the background. On top
of this background, narrow peaks will appear corresponding to the $\bar{N}N$ states
that were ``stripped" of their width by the super-radiant resonance.

\subsubsection{The $\Delta$-Nucleus system}

Several years ago an experiment on inelastic electron scattering was performed
with the reaction $^{12}$C$(e,e'p\pi^{-})^{11}$C. Evidence was found for narrow
(several MeV wide) resonances in the $\Delta_{33}$+$^{11}$C system in addition to
a broad peak of approximately 100 MeV width \cite{bartsch99}.

The free pion-nucleon resonance $\Delta_{33}$ with spin 3/2 and isospin 3/2 has
a total width of $\Gamma_{\Delta}=120$ MeV, and the centroid of the four charge
components is at 1235 MeV; this is 300 MeV above the nucleon mass. The excitation
of nuclei with various medium energy probes (electrons, protons, etc.)  reveals
a wide peak around the energy of 300 MeV. This peak is described in the framework
of $\Delta$ particle-nucleon hole ($\Delta N^{-1}$) configurations \cite{hirata79}.
The $\Delta$-component is treated as a particle bound in a potential that has
similar characteristics to the usual nucleon potential. The nuclear $\Delta$ resonance
is determined by the coupling of the $\Delta N^{-1}$ configurations to the decay
channel that involves pion emission. The $\Delta$-nucleus system has properties
that make the application of the super-radiant mechanism plausible. The $\Delta N^{-1}$
configurations are strongly coupled to a single channel $|c\rangle=|A-1;
(N+\pi)_{\Delta}\rangle$, corresponding to a (nucleon plus pion)-system with quantum
numbers of the $\Delta$ resonance and the residual $(A-1)$ nucleus. The partial
decay width of each $\Delta N^{-1}$ state is a fraction of the total width of
the free $\Delta_{33}$, of the order of a few tens of MeV. On the other hand,
the energy spacing between these states is of the order of several MeV. We are
dealing here with the case of overlapping resonances.

In Ref. \cite{AZ02} a simple shell-model description was used for the $^{12}$C
target nucleus. The nucleons occupy the lowest shell-model orbits $0s_{1/2}$
and $0p_{3/2}$ orbits. The $\Delta$ particle is assumed to move in a potential
similar to the mean nuclear field. Therefore the energy spacings between the
nuclear orbits and between the $\Delta$ orbits are close. One can then form
$\Delta N^{-1}$ states for various total spins $J$. For example, there are four
$\Delta N^{-1}$ states with $J=1^+$. Applying the theory of super-radiance we
should find three very narrow states (with widths of several MeV) and one very
broad resonance with a width around 100 MeV. The other states with spins $J=2^+$
and $3^+$ will add some narrow peaks. This picture resembles very much the observed
cross-section \cite{bartsch99}. One should bear in mind that other mechanisms might
contribute to the observed widths of these quasi-stationary states. For example,
the nucleon-hole states acquire spreading widths due to the residual nucleon-nucleon
interaction. It is important to confirm the experimental results of Ref.
\cite{bartsch99} and try to study the possible narrow structure using alternative
reactions that excite the $\Delta$-nucleus resonances.

\subsubsection{High energy phenomena}

In many applications of the super-radiant mechanism, including the last two cases,
the emphasis was placed not only on the wide super-radiant state, but also on
the fact that the mechanism of the theory creates {\sl narrow resonances} superimposed
on a background formed by the broad super-radiant state, leading in this way to
the separation of fast (direct) processes from the long-lived structures.
The consequences of this were examined and it was suggested that a mechanism
of this type can explain the existence of narrow auto-ionizing states in atoms
and narrow resonances in a number of strongly interacting hadronic systems.

In the years 2003-2004 evidence was brought for the existence of a narrow baryon
resonance with strangeness $S=+1$ around energy $E=1540$ MeV
\cite{nakano03,stepanyan03}. It was found in various experiments that the width
of this resonance (coined as $\Theta^{+}$ or $Z^{+}$), was a few MeV only.
The resonance was situated on top of a very broad (over 100 MeV) background peak.
It was difficult to understand the existence of this very narrow resonance structure
in the framework of the usual mechanism for decay of baryonic resonances, whether
exotic or non-exotic. The narrow width of the $S=+1$ exotic baryon indicates that
the decay into the kaon-nucleon continuum is quenched either by selection rules,
for example if the $\Theta^{+}$ is an isotensor, or by some dynamical features.

In view of our previous discussions, it is possible that the narrowness of the peak
observed is a result of the super-radiant mechanism \cite{AZV04}. This case serves
as a direct example of the two-level super-radiant formalism discussed in
Section 4.1. In the simplest scenario one can consider the kaon-nucleon
{\sl pentaquark} system ($s\bar{u}+ddu$ quarks) to be nonrelativistic and to form
a quasi-molecule. Molecular-like structures for non-strange pentaquarks were
discussed by Iachello \cite{iachello97}. The two particles interact via an
attractive potential of a typical range of 1 fm producing a $p$-wave resonance
at energy 100 MeV above threshold.  In Ref. \cite{jaffe03} the authors estimate that
the width of the resonance in such a potential is greater than 175 MeV, the value
typical for strong decays of baryons in this mass region. A direct calculation
gives the $p$-resonance width 190 MeV for the well of radius $a=1$ fm obtained at
the relevant energy; this required the depth of the well to be $V = 333$ MeV;
the width of 327 MeV is obtained for $a = 2$ fm and $V = 14$ MeV. In addition to
quasi-molecular states, one should consider many-quark bag dynamics in a system of
five quarks formed after the photon absorption by the original proton. Among
the intrinsic states there are groups with the same quantum numbers, including
``normal" quark states, quark-gluon states, paired states with singlet or triplet
diquark(s), states with pions and so on. If the decay width of each individual state
is greater than the energy spacing, the unperturbed spectrum will be that of overlapping
resonances. Under such conditions, the dynamics will be completely dominated by
the coupling to the decay channel, and the anti-Hermitian part of the Hamiltonian
will be the crucial factor. As a result, the super-radiant mechanism will give rise
to the observed spectrum with a very broad, $\sim$ several hundred MeV, background peak
of the super-radiant state and one or several very narrow resonances with width of
order of few MeV. For example, for two overlapping resonances with $\Delta E\approx 20$
MeV, and the bare width of $\gamma_{1,2}\approx 200$ MeV, using the equations
of Sec. 4.1, the estimate gives for the narrow resonance $\Gamma_{{\rm narr}}
\approx 4$ MeV. A more detailed and quantitative account of this can be found in
Ref. \cite{AZV04}.  In Ref. \cite{karliner04} the authors proposed somewhat different
candidates for the two interacting states. In summary, due to the overlap of unstable
intrinsic states, the super-radiant mechanism may produce the narrow peak(s) on
the broad background in exotic baryon systems with strangeness +1.

We should remark here that the observation of pentaquarks is presently quite
controversial. Some experiments that followed the initial claims of observation
have reported the absence of pentaquark signals close to energy of 1540 MeV. Also some
of the experiments that did report initially evidence for the existence of pentaquark
have now retracted these claims \cite{battaglieri06}. Nevertheless the theoretical
discussion above is still valid. The assumption of the existence of two states close
in energy around 1540 MeV might not be fulfilled. Generically, however it is possible
to obtain a very narrow width for high energy resonances when there are favorable
conditions for the super-radiant mechanism.

In the last several years, in addition to the pentaquark, a plethora of papers
were published (mainly from the $B$-factories) reporting the observation of new
resonances \cite{choi03,acosta04,abazov04} with small decay widths, much less than
typical hadrons would have at these energies (some of the reported resonances had
widths of a few MeV only). Experimental and theoretical studies suggest that among
these narrow resonances might be {\sl tetraquarks}, that is systems with two quarks
and two anti-quarks. These new resonances mostly lay around the mass region of 4 GeV,
and the proposal is that they are composed of a light quark-antiquark pair and
a heavy quark-antiquark pair, as for example $c\bar{c}u\bar{u},\;b\bar{b}u\bar{u},$
and $c\bar{c}d\bar{d}$. Heavy tetraquarks are expected to decay into heavy states
(as for example the $J/\psi$, charmonium) and light $q\bar{q}$ states, for example
pions. The decay width of such processes is expected to be large, of the order of
hundreds of MeV. However, in view of what we have discussed above, if there are
two closely spaced tetraquark states with the same quantum numbers and both are
coupled to the same final decay mode, there will be a rearrangement of widths,
one state will become very wide while the other will become narrow.

The most studied, so far, resonance among the ones suspected to be tetraquarks, is
the $X$(3872) \cite{choi03,acosta04,abazov04}, which decays into $(J/\psi)\pi^{+}\pi^{-}$.
The decay width is very small, $\Gamma< 2.3$ MeV. An unusually small width is an indication
of strong incoherence and cancelation, suggesting that the super-radiant mechanism might be
responsible for this. The broad (super-radiant) state has a very large width and
is embedded in the background. There are several observations of additional
resonances with widths not as small as that of $X$(3872), but still narrow compared
to the typical characteristic decay widths in this energy region.

The Belle collaboration reported recently the observation of the $Z$(4430) particle
\cite{choi08}. This particle is unusual because it has non-zero charge therefore
providing a strong hint that it is a tetraquark.  The dominant component of its wave
function is probably $c\bar{c}u\bar{d}$ that decays into $\psi'\pi$. The decay width
is of the order of 45 MeV. This again is an anomalously small value, especially considering
the fact that the state is located several hundred MeV above threshold. It could
possibly be viewed as a realization of the super-radiant mechanism. One should
cautiously approach the new discoveries of tetraquark states and seek more experimental
and theoretical confirmation before drawing certain conclusions about the mechanisms
responsible for narrow resonance widths. The approach outlined here presents one of possible
candidates to explain these long-lived resonances.

\newpage

\section{Quantum chaos in an open system}

\subsection{Many-body quantum chaos}

Our discussion of doorway dynamics with separation of scales corresponding
to external and internal interactions naturally leads to a general problem of
quantum chaos. The deterministic chaos in {\sl classical mechanics} is now a well
understood and discussed in textbooks \cite{schuster05} phenomenon based on
the extreme sensitivity of phase space trajectories to small variations of
initial conditions. Contrary to that, {\sl quantum chaos} \cite{gutzwiller90}
still is a subject of discussions, and some authors deny its existence
recognizing only {\sl quantum signatures of classical chaos} \cite{haake91}.
In spite of this, we can formulate generic manifestations of quantum chaotic
dynamics in a {\sl stationary} quantum system as the presence of {\sl local}
correlations and fluctuations of observables approaching the extreme
mathematical limit of canonical {\sl Gaussian ensembles} of random matrices
\cite{mehta04}.

{\sl One-body} quantum chaos emerges in systems of billiard type being caused
by the boundary conditions, for example coming from an asymmetric shape that
destroys spatial symmetries \cite{kaufman99}. Such dynamics can be successfully
reproduced experimentally with microwave cavities \cite{stoeckmann99}, especially
in the low-temperature superconducting regime with suppressed thermal dissipation
\cite{rehfeld01}. In condensed matter systems, chaotic dynamics can be generated by
the presence of disorder \cite{casati95} or external fields \cite{lerner04} and,
therefore, can be experimentally regulated.

Our main interest is in physics of {\sl many-body} quantum chaos where the chaoticity
is originated by interparticle interactions \cite{ann}. As excitation energy and level
density increase and any residual interaction mixing the independent mean-field particle
configurations becomes effectively strong, many-body quantum dynamics  in self-bound
systems approaches chaotic limit. Then the stationary states develop into exceedingly
complicated superpositions of original simple configurations. The process of chaotization
was tracked down in detail for complex atoms \cite{grib94}, nuclei \cite{big} and other
mesoscopic objects.

The ideal characteristics of quantum chaotic dynamics are given by the predictions
of the Gaussian orthogonal ensemble, GOE, (or Gaussian unitary, GUE, for the case
of violated time-reversal invariance, and Gaussian symplectic, GSE, applicable,
in particular, to time-reversal invariant systems with an odd number of fermions
and therefore Kramers degeneracy) \cite{porterbook,mehta04,guhr98,WMRMP09}. The distributions
of energy eigenvalues in these ensembles are invariant with respect to corresponding
(orthogonal, unitary, or symplectic) transformations of the basis and separated
from the distributions of the components of eigenvectors,
\begin{equation}
P_{\beta}(E_{1},\,...\,,E_{N})=C_{\beta}(N)\prod_{m<n}|E_{m}-E_{n}|^{\beta}\,
\exp\Bigl[-\beta N\sum_{n} E_{n}^{2}/a^{2}\Bigr].               \label{88}
\end{equation}
Here $C_{\beta}(N)$ are normalization constants, $\beta=1,2,4$ refer to GOE, GUE, and GSE,
respectively, while $a$ is the only available scale, a measure of the energy spread of
the whole spectrum under consideration. Typical features of such distributions are well
known: level repulsion at small spacings (linear, quadratic and quartic for those ensembles
as can be understood from elementary quantum-mechanical perturbation theory \cite{QP2}),
small probability of large gaps in the spectrum, and the maximum of the probability close
to the mean level spacing so that the levels in this limit form an aperiodic crystal with
suppressed fluctuations. Counterintuitively, spectral fluctuations are much stronger for
a regular motion when the level positions are uncorrelated and have the {\sl Poisson distribution}.

The {\sl nearest level spacing distribution} that follows from eq. (\ref{87}) is
the first (and the weakest) feature of quantum chaos, in many-body systems it is
reached already at a small interaction strength \cite{big}. More sensitive measures
deal with the fluctuations of the level number along large fragments of the spectrum
\cite{Delta3}. Such an analysis is useful for checking the completeness and purity of
long empirical level sequences, for example for low-energy neutron resonances which
are important in practical applications \cite{undraa03,mulhall07,mulhall09}.
For the purpose of this review, we have to emphasize that such statistical applications
assume that the long-lived resonances can be considered as representatives for
the sequences of {\sl stationary} states although in reality they belong to
the {\sl continuum}.

The most interesting and far-reaching predictions of many-body quantum chaos refer to
the structure of stationary states in the chaotic regime. Strongly mixed states in
a narrow energy window have essentially the same degree of complexity \cite{percival}
and the same macroscopic properties corresponding to thermodynamic equilibrium which
provides a modern justification of statistical mechanics \cite{ann}. In the limit of
random matrix theory, for large dimensions of the Hamiltonian matrices, the components
of the chaotic wave functions are uncorrelated Gaussian variables. Consider, for example,
an experiment that singles out a certain component of the wave function. Probing various
states of similar structure, one would then discover that the distribution of corresponding
probabilities is the chi-square for one degree of freedom, called the {\sl Porter-Thomas
distribution} (PTD) in nuclear spectroscopy \cite{BM}. The old experiments for neutron
resonances confirmed this prediction. The joint analysis of $\nu>1$ components, in
a similar way, is expected to reveal the chi-square distribution for $\nu$ degrees
of freedom.

The closer study of chaotic wave functions shows \cite{SF} that small perturbations acting
in this region of states turn out to be enhanced compared to their effects for regular wave
functions of simple configurations. This follows from the fact that, on the road to chaos,
the matrix elements of simple operators between chaotic functions of a similar degree of
complexity decrease slower than the energy denominators in perturbation
theory for the high level density. This {\sl chaotic enhancement} is responsible for
the unusually high, up to 10\%, level of the parity violation effects in interactions of
slow neutrons with heavy nuclei \cite{auerbach92,bowman93}. The chaotic mechanism of the enhancement is confirmed by the experiments on parity violation in nuclear fission by longitudinally polarized neutrons \cite{kotzle}. The significant asymmetry of heavy
fragments with respect to the neutron spin does not depend on the final distributions
of product masses and kinetic energies being predetermined in the chaotic (``hot") stage
of the compound nucleus.

Among other practically important applications of the ideas of quantum chaos we can
mention the extraction of information about invisible fine structure states in experiments
with insufficient resolution \cite{kilgus87}. New computational methods for large-scale diagonalization of Hamiltonian matrices of extremely high dimension are based on the
chaotic properties of highly excited configurations \cite{expconvHVZ,expconvHBZ} and
resulting {\sl exponential convergence} of the quantities of spectroscopic interest $-$ low-lying energies and other observables $-$ as a function of progressive truncation
of the matrix preordered in a specific way.

\subsection{Statistics of resonances}

Many experimentally known sequences of complicated quantum states with identical global
symmetry in fact are resonances observed in the continuum spectrum and therefore having
finite lifetime. A natural question is if the properties of quantum chaos predicted by
the random matrix theory for closed systems are still valid for quasistationary states.

Quantum theory of resonance states in complex systems is mostly developed for two limits
clearly distinguished, for example, in neutron reactions on heavy targets. Not very far
from thresholds, the neutron resonances appear as quite narrow well isolated peaks in
neutron cross sections. Each resonance here corresponds to a long-lived compound state
probed in a specific mode of its excitation (entrance channel) and decay (exit channel).
In elastic neutron scattering, the experiment singles out a certain component of the quasistationary wave function, namely the one corresponding to the continuum neutron
(typically in $s$-wave at low energies) and the ground state of the target. Here the
resonance neutron widths, being proportional to the weight of this component in the
compound state, are expected to obey the PTD, as for genuinely stationary states.
In the opposite limit of broad resonances, their widths exceed the energy spacings.
The cross sections in the region of overlapping resonances manifest a complex picture
of wild interference ({\sl Ericson fluctuations} \cite{ericsonAP63,EMK66}) with the correlational length of cross sections that is usually assumed to be determined by
the average width of overlapping resonances. The latter quantity is defined by
the inverse mean lifetime of complicated intrinsic states. The deviations from this
limiting picture are known experimentally for a long time, see for example
\cite{mitchell86,kanter78,rotter91} and references therein. The extreme statistical
averaging which is a core of the Ericson theory does not account for unitarity
requirements which imply collective super-radiance and trapping phenomena occuring
even if internal dynamics is chaotic.

%
%\subsubsection{Isolated resonances (weak continuum coupling)}

In the no-overlap regime of resonances in a chaotic many-body system, one can still
expect the statistics of resonance parameters (energies and widths) to be in
approximate agreement with the predictions of random matrix theory formulated for
stationary states. This opinion was widely accepted but the recent accumulation of
experimental data of higher precision \cite{koehler10,nature} put this statement
in doubt. Indeed, we mentioned, eq. (\ref{88}), the stationary level repulsion
at small distances. However, this is a property of a mixing by a {\sl Hermitian}
interaction. For unstable states, their energies are defined within uncertainty
related to their lifetime $\tau\sim\hbar/\Gamma$. There is no reason to expect
the repulsion of resonances at short distances in energy comparable with their
widths. The exact description in terms of the effective {\sl non-Hermitian}
Hamiltonian confirms the disappearance of such repulsion as can be seen already
from our two-level model above, Sec. 4.1. {\bf Fig. 6} taken from Ref.
\cite{mizutori93}, where the case of a single open channel was studied, shows that,
as a function of the increasing continuum coupling constant $\kappa$, the probability
$P(0)$ to have coinciding or very close positions of resonances on a real energy
axis (level crossing) is growing from zero and falls down again after
the super-radiant transition when the broad state spreads into a part of the background.
The dependence on the number of open channels and strength of the intrinsic interaction
is illustrated by {\bf Fig. 7} \cite{celardo08}. In the GOE case all curves are
qualitatively similar but the effect is getting more pronounced with the number
of channels increasing. The model with two-body random interactions (TBRE),
see also Section 7.4,  shows that after the super-radiant transition $P(0)$ falls
down as a function of $\kappa$ in a universal way since, independently of the strength
of the two-body interaction, we are dealing with uniformly trapped states.

Moreover, the same qualitative argument of the preceding paragraph predicts also that
energy centroids and widths of resonances cannot be independent, they have to be
correlated, $-$ the phenomenon clearly outside the standard theory. We need to
construct a new statistical ensemble for unstable states generalizing the canonical
Gaussian ensembles for closed systems. The similar ensemble of complex matrices
\cite{ginibre65} cannot serve as an appropriate model since its eigenvalues are spread
over the complex plane while the physical widths correspond to the {\sl negative}
imaginary parts of the poles of the scattering matrix $-$ resonance widths $\Gamma_{r}$
are positive. This property is guaranteed by the construction of the effective
Hamiltonian ${\cal H}$ [the choice of $E^{(+)}$ in eq. (\ref{24})]. The simplest
appropriate statistical ensemble for unstable states (our ${\cal Q}$-space) assumes
the intrinsic GOE dynamics ($H_{{\cal QQ}}$ part) along with uncorrelated decay
amplitudes (parameters $A^{c}_{1}$ of the coupling Hamiltonian $H_{{\cal PQ}}$).
These amplitudes include kinematic factors depending on energy and vanishing at
thresholds. As mentioned in our discussion of the continuum shell model, the energy
dependence is crucial in realistic shell model studies. Dividing out this dependence
we come to the {\sl reduced widths} which can be considered as statistically
independent. The PTD was expected to work just for these reduced quantities.

Assuming the statistical independence and similarity of all decay amplitudes one can
take them as random Gaussian quantities with zero mean and variance
\begin{equation}
\overline{A^{c}_{1}A^{c'}_{2}}=\delta_{12}\delta^{cc'}\,\frac{\gamma^{c}}{N},     \label{89}
\end{equation}
where $N$ is a large dimension of the ${\cal Q}$-space, and $\gamma^{c}$ parameterize characteristic partial widths. The ansatz (\ref{89}) is natural for the system with
chaotic internal dynamics. However, one can argue that, in realistic applications,
the assumption (\ref{89}) can be reasonable, independently of the presence of internal
chaos, like in the case of weakly chaotic or even regular intrinsic dynamics, for
$N\gg 1$. In self-bound rotationally invariant systems, such as a heavy nucleus,
we need to consider the states with the same quantum numbers $J,M$ of angular momentum
(plus parity and isospin). For a given shell-model partition, there are many ways of
angular momentum coupling leading to the same total symmetry. These paths expressed
through the fractional parentage coefficients are essentially chaotic. The idea of
random walk in the angular momentum space was used long ago by Bethe \cite{bethe37}
for estimates of the spin-dependent nuclear level density. This {\sl geometric
chaoticity} plays an important role in statistics of self-bound systems
\cite{MVZ,expconvHVZ,expconvHBZ,papenweid06} and it can justify the assumption
(\ref{89}) even for the case of intrinsic interactions which are relatively weak to
introduce the full dynamical chaos. Irrespectively of dynamic interaction, this leads
to chaotic structure of wave functions and, for example, to the predominance of the
scalar representation $J=0$ in the ground state of an even-even nucleus after averaging
over all interactions allowed by the conservation laws \cite{johnson98,VZmeso}.

With the GOE intrinsic dynamics and a {\sl single decay channel}, it was derived
algebraically \cite{SZPL88,SZNPA89} that the common distribution function of complex
energies (\ref{6}) is given by the Ullah distribution \cite{ullah69},
\begin{equation}
P({\cal E}_{1},\,...\,,{\cal E}_{N})=C_{N}\left(\prod_{n}\,\frac{1}{\sqrt{\Gamma_{n}}}
\right)\,\left(\prod_{m<n}\,\frac{|{\cal E}_{m}-{\cal E}_{n}|^{2}}{|{\cal E}_{m}-
{\cal E}^{\ast}_{n}|}\right)\,\exp[-NF(E_{n},\Gamma_{n})],            \label{90}
\end{equation}
where the ``free energy" of interacting resonances is given by
\begin{equation}
F=\,\frac{1}{a^{2}}\,\sum_{n}E_{n}^{2}\,+\,\frac{1}{\gamma}\sum_{n}\Gamma_{n}\,+
\,\frac{1}{2a^{2}}\,\sum_{m<n}\Gamma_{m}\Gamma_{n}.                   \label{91}
\end{equation}

The distribution in the complex energy plane (\ref{90}) reflects the interplay of
various dynamical factors. The product of inverse square roots of the widths is
the remnant of the PTD that is produced by the transformation from the amplitudes
of given components in a complicated superposition of wave functions to their
probabilities. The next product has the numerator responsible for the quadratic
repulsion of energies {\sl in the complex plane}, similar to the Gaussian unitary
ensemble because the openness of the system effectively works as violation of
time-reversal invariance. The denominator in this product shows ``attraction" of
the widths to their ``electrostatic images" and therefore the accumulation of
the widths in the vicinity of the real axis (many long-lived states). The real
axis remains a singular boundary that excludes the appearance of negative widths.
We can recall that the electrostatic analogy was fruitful in the first studies of
quantum chaos by Dyson \cite{dysone-st}. The free energy (\ref{91}) is the main
factor regulating the equilibrium arrangement of the complex energies.
The super-radiance is contained in the last sum (repulsive width interaction).
When typical widths grow beyond average spacings of real energies, the last term in $F$ suppresses the probabilities of all configurations with uniform width distribution.
The prominent maximum of probability is realized then on the configuration with one
predominant width (super-radiant state). This means that the effect of super-radiance
may coexist with chaotic internal dynamics.

In {\sl many-channel} situations, direct algebraic derivation of the joint distribution
function for complex energies is difficult and the more advanced methods related to supersymmetry should be applied. Such methods were developed first for closed systems \cite{efetov} and then applied to unstable systems as well \cite{supersym}. Unfortunately,
the results are typically expressed in the form of cumbersome multi-dimensional integrals
that makes the analysis not an easy task (the calculations are usually simpler for
the intrinsic part described by the GUE rather than GOE). In this situation the direct
numerical simulation of statistical ensembles becomes the method of choice. We have
seen in {\bf Fig. 2} \cite{izr94} the evolution of the distribution of complex energies
as the strength of continuum coupling grows. The effect of super-radiance is clearly
revealed as the segregation of super-radiant states (in the amount equal to the number
of open channels) from the larger cloud of trapped states. This transformation is
rather sharp and reminds a phase transition. Regrettably, the experimental data for multi-resonance systems with adjustable parameters that would allow to trace this
restructuring are practically absent. The best candidates for this purpose are probably experiments with superconducting microwave cavities \cite{rehfeld01} and with acoustic
chaos \cite{acchaos}.

\subsection{Statistics of resonance widths}

The statistical distribution of the resonance widths is an essential characteristic of
a sufficiently long sequence of resonances. The low-energy neutron scattering off heavy
nuclei provides rich statistics of such data. Usually the primary treatment of the data
in the regime of well separated resonances is accomplished with the aid of the $R$-matrix analysis \cite{descouvemont10} which we do not discuss here. At energies below the first inelastic threshold, neutron scattering can be only elastic, and we can apply the
one-channel consideration. We will also omit a discussion of extracting the energy
dependence that leads to the reduced widths. In practice this procedure should depend
on the specific features of the compound system since the standard energy dependence
defined by the orbital momentum of the neutron may be distorted by the shape of the single-particle strength function in the closed system \cite{weidenmueller10}. Assuming
that we can have a reliable set of reduced resonance parameters, we can apply
the statistical measures.

For the set of physically equivalent reduced widths of a single-channel reaction one
could expect the PTD. The admixture of intruder resonances corresponding to different
quantum numbers, such as $p$-wave resonances in the predominantly $s$-wave sequences,
should violate the level spacing distribution and statistics of level fluctuations.
Such an impurity supposedly will move the distribution in the direction of the two-channel situation, and therefore {\sl increase} the effective number of degrees of freedom given
by the parameter $\nu$ in the chi-square distribution,
\begin{equation}
P_{\nu}(\gamma)=C_{\nu}\gamma^{\nu/2-1}e^{-\gamma/\bar{\gamma}}.       \label{92}
\end{equation}
Here we can notice that even for well studied cases, such as $^{235}$U, the undetected
admixture of $p$-wave can reach few percent \cite{mulhall07}. Also we select an even-even target, so that all $s$-wave resonances have the same spin 1/2. For many years, based
on experimental data, it was thought that the PTD, eq. (\ref{92}) with $\nu=1$, describes
well the reduced neutron width distributions. However, recent precise measurements \cite{koehler10,nature} claim that there are statistically reliable deviations from
the PTD in the direction of the effective parameter $\nu$ {\sl smaller than one}.

The proper inclusion of the effects of coupling to continuum \cite{luca10} with the use
of the effective non-Hermitian Hamiltonian shows that such deviations inevitably appear
even for the sequence of statistically equivalent resonances although complementary
effects of single-particle strength functions \cite{weidenmueller10} can also be
important. The PTD corresponds to the complete factorization of the joint distribution (\ref{87}) into the independent GOE-like distribution of the resonance centroids $E_{n}$
and the distribution of non-interacting widths. As seen from (\ref{86}) this occurs only
in the extreme limit of vanishingly small widths $\Gamma_{n}$ in comparison to the energy spacing $D\sim a/N$. The energy-width correlations enter the game and lead to the deviations from the PTD already at rather small value of the coupling strength $\kappa\sim \gamma/D$.
The evolution of the width distribution as a function of increasing continuum coupling
is illustrated by {\bf Fig. 8} \cite{luca10}. Very soon, the chi-square model
itself becomes a very bad approximation to the distribution predicted by (\ref{90}). In agreement with the random matrix results \cite{fyodorov97,sommers99}, the tail of the distributions behaves $\propto (\Gamma/\overline{\Gamma})^{-2}$ that cannot be described
by chi-square distributions.

\subsection{Role of intrinsic dynamics}

Another interesting aspect of the width distribution problem is related to the intrinsic dynamics. The exact results cited above all refer to the canonical Gaussian ensembles
of random matrices. In such cases of developed quantum chaos, all intrinsic states are
of the same degree of complexity and their uncorrelated components give rise to Gaussian distributions, essentially as a manifestation of the central limit theorem. A slightly
different type of statistics arises if the interparticle interactions, being still random,
carry certain selection rules. We have already mentioned the role of geometric chaoticity
in the process leading to complicated mixing of the nuclear wave functions even in
the absence of interactions.

{\sl Two-body interactions} being the most relevant part of intrinsic dynamics bring
a regular element in the Hamiltonian matrix of a closed system: many zeros and many
repeated substructures \cite{flizr97}. Therefore, even if the resulting local pattern
of the level repulsion and spacing distribution is in this case similar to the GOE limit,
the remaining correlations may have their influence upon the fine features, including
the continuum coupling. This can be modeled with simulating the intrinsic Hamiltonian
$H_{{\cal QQ}}$ by the {\sl two-body random ensemble} (TBRE). Here the two-body matrix
elements are uncorrelated random quantities with zero mean and variance $v^{2}$.
The ratio $v/d$ of the TBRE strength to the appropriate mean-field level spacing
characterizes the relative strength of intrinsic interactions. The estimates of
the critical strength for the onset of chaos in the Fermi system of shell-model type
with a certain number of particles in a truncated orbital space were given in Ref. \cite{flizr97a}.

{\bf Fig. 8} shows the best value for the chi-square parameter $\nu$ for different
values of the continuum coupling strength $\kappa$ found \cite{luca10} in the simulation
of the dynamics by the GOE and TBRE with and without strong intrinsic interaction.
The deviations from the chi-square distribution grow substantially faster for more
regular dynamics both for one and two open channels. Intrinsic chaos is accompanied by
the level repulsion and their relative ordering. Absence of chaos implies a stronger
sensitivity to the continuum coupling. This physics sets the limits for the applicability
of the traditional interpretation of the width of a neutron resonance as a measure for
the strength of a pure neutron component in the wave function of compound nucleus.
At the same time, we can acquire an instrument of evaluating the degree of intrinsic
chaos by the observable width distributions for sufficiently clean resonance sequences.
In general, various studies allows one to conclude that increasing the intrinsic
interaction we suppress the fluctuations in the continuum making the compound states
more uniform.

\newpage

\section{Signal transmission through a mesoscopic system}

\subsection{Common physics of mesoscopic systems}

Scattering of an external beam off a many-body system is the most powerful instrument for obtaining
the indispensable information on the internal structure of the system. At the same time, the situation
can be inverted and our goal might be just the propagation of the signal itself \cite{mello04}. The signal can
be not only transmitted - we may need to transform it, split and recombine, separate resonance components,
selectively amplify or suppress parts of the signal etc. It might be important to implement such different
tasks simultaneously and record the results. All these tasks are necessary for quantum computing and in
all modern applications of information treatment. A parallel problem is that of finding or synthesizing
the quantum mesoscopic medium especially appropriate for a specific purpose.

In fact, all these problems can be addressed in the framework of the effective non-Hermitian Hamiltonian.
Here we have the intrinsic system with possible resonances along with open exit and entrance channels,
plus some channels closed at a certain energy (frequency) of the signal but virtually influencing
the results of a process. A system can also be embedded into complex environment and subject to external
noises and decoherence but this set of problems is beyond the scope of the current review although
it can be analyzed with similar tools.

The terminology borrowed from nuclear reactions as a fully developed subfield can be successfully
used for other mesoscopic systems. In all cases of the signal transmission we are interested in
the exit-entrance relation given by the scattering $S$-matrix in the space of open channels.
This matrix describes the probabilistic connection between the in- and out- quantum states.
Without restriction of generality, this unitary matrix is defined by eq. (\ref{38}), where we will
be interested mainly in the second part that describes the signal propagation inside the system
(compound states). The observable quantities are {\sl cross sections, or transmission probabilities,
their correlations and fluctuations} as functions of energy and intrinsic parameters of the system.

\subsection{Statistics of cross sections}

If the signal lives inside the system for sufficiently long time (longer than the Weisskopf time $\hbar/D$,
where $1/D$ is the typical density of states coupled to the signal), the situation becomes similar
to the compound nucleus. In the experiment with good resolution we expect the cross section to reveal
isolated narrow resonances corresponding to the levels of the compound system. As lifetime of intrinsic
states shortens, the resonances broaden and start to overlap. Here one typically expects strong fluctuations
of observable cross sections or transmission. The classical Ericson theory for the region of overlapping
resonances \cite{ericsonAP63,EMK66} does not analyze the properties of individual resonances, concentrating
instead on the observable statistical features. However, this theory does not account for the super-radiance
effects since the coupling of intrinsic states through the continuum is not included into consideration.

If we neglect direct reactions, the $\hat{S}$-matrix can be presented in an explicitly unitary form (\ref{38}),
\begin{equation}
\hat{S}=1-i\hat{T}=\,\frac{1-(i/2)\hat{K}}{1+(i/2)\hat{K}},                                      \label{93}
\end{equation}
where the $K$-matrix (\ref{35}) is given by the Green function of intrinsic propagation sandwiched between
the continuum amplitudes,
\begin{equation}
\hat{K}=\,{\bf A}\,\frac{1}{E-H}\,{\bf A}^{T}.                          \label{94}
\end{equation}
The scattering characteristics can be subdivided into average and fluctuational parts \cite{fridman}.
For example, it is assumed for the scattering amplitude that
\begin{equation}
T^{ab}(E)=\,\overline{T^{ab}(E)}\,+\,T^{ab}_{{\rm fl}}(E),                  \label{95}
\end{equation}
where the averaging is performed over various resonances or, in practice, over an energy interval
including sufficiently many resonances of supposedly similar nature. The same separation of contributions
is performed for the $S$-matrix and cross sections.

In the absence of direct reactions avoiding formation of the compound system, only elastic channels contribute
to the average amplitude, $\overline{T^{ab}(E)}\propto\delta^{ab}$. In the language of the effective Hamiltonian,
this is equivalent to the discussed above statistical ansatz (\ref{89}) for the coupling amplitudes.
Under this assumption, the averaging of the $\hat{K}$-matrix gives \cite{celardo07}
\begin{equation}
\overline{K^{ab}(E)}=-i\delta^{ab}\kappa^{a}, \quad \kappa^{a}=\,\frac{\pi\gamma^{a}}{ND}, \label{96}
\end{equation}
where the density of states $\rho=1/D$ should be taken in the center of the averaging energy interval
containing $N$ resonances. At $N\gg 1$, this leads to the diagonal average $\hat{S}$-matrix,
\begin{equation}
\overline{S^{ab}}=\delta^{ab}\,\frac{1-\kappa^{a}}{1+\kappa^{a}}.                  \label{97}
\end{equation}
Assuming equivalent channels, $\kappa^{a}=\kappa$, we can express the total cross section with the aid
of the optical theorem,
\begin{equation}
\overline{\sigma_{{\rm tot}}}=\sum_{b}\overline{\sigma^{ba}}=2\Bigl(1-{\rm Re}\overline{S}\Bigr)=
\frac{4\kappa}{1+\kappa},                                                        \label{98}
\end{equation}
where we omit the prefactor $4\pi/k^{2}$.

It is convenient to work with the {\sl transmission coefficient} for each channel,
\begin{equation}
\tau^{a}=1-|\overline{S^{aa}}|^{2}=\,\frac{4\kappa^{a}}{(1+\kappa^{a})^{2}}.   \label{99}
\end{equation}
According to the definition (\ref{96}), this quantity, as well the average characteristics like
(\ref{98}), does not depend on the intrinsic dynamics if the parameter $\kappa$ is always
renormalized for the proper level density. The so-called {\sl perfect coupling} regime in channel
$a$, $\tau^{a}=1$, corresponds to $\kappa^{a}=1$ when the average $\hat{S}$-matrix (\ref{96}) vanishes.
This is precisely our region of the super-radiant transition. We can note that the transmission
(\ref{98}) is symmetric with respect to the substitution $\kappa\rightarrow 1/\kappa$;
in the language of super-radiance this means that at large $\kappa$ one of $N\gg 1$ resonances
broadens and disappears, and the transmission pattern in a given channel returns to the regime
of no overlap.

Similarly, the behavior of the fluctuating cross sections in various (by assumption, equivalent)
channels can be encoded into another important parameter, the {\sl elastic enhancement factor},
the ratio
\begin{equation}
F=\,\frac{\overline{\sigma_{{\rm fl}}^{aa}}}{\overline{\sigma_{{\rm fl}}^{ba}}}, \quad b\neq a,
                                                                     \label{100}
\end{equation}
of elastic to inelastic cross sections \cite{harney86}. In terms of the transmission coefficient
and the elastic enhancement factor, we can derive
\begin{equation}
\overline{\sigma^{{\rm el}}}=F\overline{\sigma^{{\rm inel}}}=\,\frac{F\tau}{F+M-1}.
                                                                    \label{101}
\end{equation}
The conventional Hauser-Feshbach parametrization with $\overline{\sigma^{ab}}
\propto (1+\delta^{ab})$ leads to $F=2$. This result was repeatedly derived in
various approaches \cite{agassi75,supersym,verbaarschot86} in the regime of
strong intrinsic chaos. The numerical analysis \cite{celardo07} shows that
the elastic enhancement factor goes down from $F\approx 3$ to $F=2$ with increase
of the interaction strength reaching the standard value $F=2$ in the GOE limit.
This behavior requires more detailed understanding. {\bf Fig. 9} from Ref.
\cite{celardo07} shows the average elastic fluctuational cross section in
the TBRE simulation close to the critical coupling regime, $\kappa=0.8$,
as a function of the intrinsic TBRE interaction strength $\lambda=v/d$.
We see the sharp transition to the GOE limit at the same value of $\lambda$
that corresponds to onset of chaos in the closed system.

\subsection{Fluctuations of cross sections}

It is accepted by the standard Ericson theory that in the overlapping regime
the variance of the width distribution is small compared to the average width.
Essentially this is based on the idea of partial widths being independent random
variables. From our consideration of super-radiance it is clear that there is no
room for Ericson fluctuations in the one-channel case because of a sharp transition
from isolated resonances to the situation of one broad and many trapped states
avoiding the overlapping regime. For many channels, $M\gg 1$, one could
expect that the partial widths are independent random variables, so that both,
the variance Var$(\Gamma)$ and the average total width $\overline{\Gamma}$,
grow linearly with $M$. Then the ratio $g={\rm Var}(\Gamma)/
\overline{\Gamma}^{2}$ falls down as $1/M$. We have discussed earlier
the width distribution for one open channel at realistic (small for the neutron
resonances) values of $\kappa$. At $\kappa\sim 1$, the ratio $g$ is not small
and strongly depends on $\kappa$. As a function of $M$, {\bf Fig. 10},
it falls much slower than $1/M$ revealing a trend to saturation; one can
clearly see the dependence on intrinsic dynamics (parameter $\lambda$) that is
getting stronger as $\kappa$ grows. Again the onset of intrinsic chaos is
smoothing the width fluctuations.

In the case of intrinsic dynamics described by the GOE, the simulations agree
with the prediction of a singularity at the critical coupling,
\begin{equation}
g=\,\frac{{\rm Var}(\Gamma)}{\overline{\Gamma}^{2}}\,\propto\,\frac{\ln^{2}
(1-\kappa)}{(1-\kappa)^{2}}.                                             \label{102}
\end{equation}
This singularity is revealed, for any value of $\lambda$, in the TBRE case as well.
Another characteristic feature is the presence of correlations between the widths
and the fluctuating part of the resonance amplitudes of the $S$-matrix.
Such correlations are amplified in the region of critical coupling.

The conventional theory of overlapping resonances predicts that the correlation
function of cross sections,
\begin{equation}
C_{\sigma}(\epsilon)=\,\overline{\sigma(E)\sigma(E+\epsilon)}\,-\,\overline{\sigma(E)}^{2},
                                                                     \label{103}
\end{equation}
as well as the similar correlation function $C_{T}(\epsilon)$ for the scattering amplitude are
essentially the same. Being expressed as ratios $r(\epsilon)=C(\epsilon)/C(0)$, they have
a simple form
\begin{equation}
r(\epsilon)=\,\frac{l^{2}}{\epsilon^{2}+l^{2}},                   \label{104}
\end{equation}
and the correlation lengths coincide, $l_{\sigma}=l_{T}$, being equal to the average width
$\overline{\Gamma}$. The qualitative argument in favor of this identification uses the idea that
it is possible to keep correlations only on a time scale of the mean lifetime of
the resonances, $\sim\hbar/\overline{\Gamma}$. Such arguments do not account for the effects
of super-radiance. The analysis \cite{celardo07} confirmed that the two correlation lengths
are equal for large $M$, while $l_{\sigma}<l_{T}$ for small number of channels, when
the Ericson assumption are invalid, especially for weak intrinsic interaction. The fluctuations
are not reproducible in different channels of the reaction. The correlation
function is also different from eq. (\ref{104}), in agreement with Ref. \cite{dittes92}.
The equality $l=\overline{\Gamma}$ is not valid, except for the case of very small $\kappa$,
as it was recognized long ago \cite{kanter78,brody81,moldauer75}. Indeed, in the presence of the
super-radiant trend, there is no justification for considering these quantities identical.
Instead, the correlation length at large $M$ agrees well with the expression
\cite{lehmann95} through the transmission coefficient,
\begin{equation}
\frac{l}{D}\,=\,\frac{M\tau}{2\pi}\,= \frac{M}{2\pi}\,\frac{4\kappa}{(1+\kappa)^{2}}.
                                                                   \label{105}
\end{equation}
Here the Weisskopf time $\hbar/D$ plays the role of the characteristic unit to measure the
intrinsic life of the signal inside the system. For the same level density $1/D$, the results
do not depend on the intrinsic interaction strength $\lambda$. The general relation between
the average resonance width and the transmission coefficient is given by the Moldauer-Simmonius
formula \cite{moldauer67,simmonius74},
\begin{equation}
\frac{\overline{\Gamma}}{D}=-\,\frac{M}{2\pi}\,\ln(1-\tau).    \label{106}
\end{equation}
Only at small $\tau$ eqs. (\ref{105}) and (\ref{106}) lead to the equality $l=\overline{\Gamma}$.

\subsection{From Ericson to conductance fluctuations}

The so-called {\sl universal conductance fluctuations} \cite{altshuler85,lee85,beenakker97,imry99}
in electronic mesoscopic systems can be considered on the same footing as nuclear
reactions; this analogy was repeatedly discussed in the literature
\cite{weidenmueller90,celardo07,celardo08}. The analysis with the aid of the effective
non-Hermitian Hamiltonian \cite{sorathia09} demonstrates how even delicate predictions
of theory of universal conductance fluctuations are reproduced by the consideration based on
a general quantum-mechanical foundation.

The {\sl conductance} of a mesoscopic system can be defined according to Landauer
as \cite{beenakker97} a sum of partial cross sections,
\begin{equation}
G=\sum_{b=1}^{M/2}\,\sum_{a=M/2+1}^{M}\sigma^{ab}.    \label{107}
\end{equation}
In order to establish the direct connection with conductance theory, we divide here
our $M$ channels, $M$ considered to be an even number, into two equal groups
of entrance, $b$, and exit, $a$, channels (this limitation certainly can be eliminated, see
for example \cite{botten07}); we also omit the standard dimensional factor
$2e^{2}/\hbar$. For equivalent channels, the average conductance can be directly expressed
through the average cross section (\ref{100}) and the elastic enhancement factor,
\begin{equation}
\overline{G}=\left(\frac{M}{2}\right)^{2}\overline{\sigma^{ab}}=
\frac{M^{2}\tau}{4(F+M -1)}.                       \label{108}
\end{equation}
At $M\gg 1$, the limiting value $M\tau/4$ is fully determined by the number
of channels and the transmission $\tau$. The simulations \cite{sorathia09} are in good
agreement with this prediction both for intrinsic interactions described by the GOE and
by the TBRE.

The fluctuations of conductance depend on the degree of chaos inside the system.
The analytical results available for the GOE intrinsic dynamics
\cite{guhr98,mello04,beenakker97} predict the variance Var$(G)$ evolving between 2/15 in
the ballistic regime to 1/8 in the diffusive regime. The approach of the effective
Hamiltonian indeed reproduces these values, see {\bf Fig. 11} taken from \cite{sorathia09}.
The crossover from one to another regime occurs precisely at the value
of the intrinsic interaction parameter $\lambda$ that corresponds to the onset
of chaos in the closed system decoupled from the continuum. This and more detailed
results on Var$(G)$, both for the GOE and for other models of internal dynamics,
can be used for extracting information on the degree of chaos inside the system
from the experimental conductance data, for example, as a function of the degree
of openness of quantum dots. Especially useful might be results on the various
correlations of ``elastic" ($a=b$) and ``inelastic" $(a\neq b)$ cross sections;
the correlations turn out to be negative for the processes without a common channel.
As a function of the continuum coupling, the maximum sensitivity to the intrinsic chaos
is observed at the intermediate strength of continuum coupling while at the perfect
coupling the continuum effects overshadow the information coming from inside.
An interesting object of application of these results can be found in correlations
of speckle patterns \cite{feng88,freud88}.
%Fig.4\sorathia09

\subsection{Transport through nanostructures}

Without trying to give a full review of this extremely rapidly developing area,
we only point out the relation of this tempestuous activity to the main topic
of this article. In Section 2.2 we gave an example of the tight-binding model
for one-dimensional lattice serving as a quantum wire for transmission of a signal,
in this case propagation of an electron through a sequence of potential wells
coupled by tunneling among themselves and through end barriers to the external
world. This simplified model discussed originally in Refs. \cite{SZAP92} and
\cite{VZWNMP04} was recently generalized and put on a realistic foundation
\cite{celardo09,PRB10}. The whole approach with the idea of super-radiance and
trapping in an open many-body system described by the effective non-Hermitian
Hamiltonian was translated into the language of quantum dots in Ref. \cite{berkovits04}.
This is a prototype of many practical arrangements including semiconductor
superlattices and quantum dot arrays \cite{tsu73,pacher01,morozov02}, molecular
bridges \cite{walter04,maiti07} etc. Similar physics turns out to be appropriate
for the description of energy transfer in biological ``antennae" (like photosynthetic
bacteria \cite{ray99}). The external barriers implement the controllable coupling
to the continuum. It was shown \cite{celardo09} that a realistic sequence of potential
barriers can be precisely modeled in the same approach and related to the open Anderson
model \cite{anderson58} (the exact correspondence holds for weak coupling between
the sites).

A general situation with {\sl asymmetric coupling}, recall the parameters
$\gamma_{L}$ and $\gamma_{R}$ in the Hamiltonian (\ref{7}) and the solution
for the resonances given by eqs. (\ref{12}) and (\ref{13}), was analyzed in detail
in Ref. \cite{PRB10}. Two realistically important cases are distinguished by
the ordered character of the chain in contrast to the presence of internal
disorder. In the case of a chain with $N\geq 2$ sites without disorder,
the transport near the center of the energy band (\ref{8}) reveals perfect
transmission $\tau=1$ at all resonance peaks for the critical value,
$\gamma_{{\rm cr}}=2v\sqrt{q}$ of the entrance width, where $v$ is the tunneling
amplitude, and we set $\gamma_{L}=\gamma,\,\gamma_{R}=\gamma/q$ introducing
the asymmetry parameter $q$. The same value $\gamma_{{\rm cr}}$ provides the maximum
of the transmission integrated over the energy band.

The transmission is determined by the sums (\ref{13}) over the Bloch states inside
the energy band. These sums can be calculated \cite{PRB10} as functions of
$\epsilon=E/2v$ in the closed form,
\begin{equation}
P_{\pm}(\epsilon)=(1\pm 1)\,\frac{\epsilon}{4v}-\,\frac{\epsilon^{2}-1}{v(z_{+}-z_{-})}
\left[\frac{z_{+}^{\alpha_{\pm}}} {z_{+}^{2N+2}-1}-\,\frac{z_{-}^{\alpha_{\pm}}}
{z_{-}^{2N+2}-1}\right],                                     \label{109}
\end{equation}
where $z_{\pm}=\epsilon\pm\sqrt{\epsilon^{2}-1},\;\alpha_{+}=2N+2,$ and $\alpha_{-}=N+1$.
Inside the band $\epsilon=\cos\beta$ and $z_{\pm}=\exp(\pm i\beta)$. This leads to
the ``elastic" transmission (in this formulation back scattering) expressed in terms of
$\eta_{L.R}=\gamma_{L,R}/2v$,
\begin{equation}
\tau^{LL}=4\eta_{L}^{2}\,\frac{\eta_{L}(1+\eta_{L}\eta_{R}+\eta_{R}^{2})+
\eta_{R}\cos(2\beta)+(1+\eta_{R}^{2})
\sin\beta}{(\eta_{L}+\eta_{R})(1+\eta_{L}\eta_{R})(1+\eta_{L}^{2}+2\eta_{L}\sin\beta)},
                                                          \label{110}
\end{equation}
and analogously for $\tau^{RR}$. The "inelastic" channel (transmission) is characterized by
\begin{equation}
\tau^{LR}=4\eta_{L}\eta_{R}\,\frac{\sin\beta}{(\eta_{L}+\eta_{R})(1+\eta_{L}\eta_{R})},
                                                          \label{111}
\end{equation}
so that the condition of unitarity, $\tau^{LL}+\tau^{LR}=-2\,{\tt Im}\,T^{LL}$ is
fulfilled. The transmission integrated over the energy band of the width $4v$ and
shown in {\bf Fig. 12} is well described by the same equation that is valid for two intrinsic
states,
\begin{equation}
\frac{1}{4v}\,\int dE\,\tau(E)=\,\frac{\pi\gamma}{2v}\,\frac{1}{(q+1)[1+\gamma^{2}/(4v^{2}q)]}.
                                                         \label{112}
\end{equation}
The transmission maximum corresponds to $\gamma=\gamma_{{\rm cr}}$, or $\gamma_{L}\gamma_{R}=
4v^{2}$, the point of the ``balance" between decay outside and tunneling inside
the chain. This condition agrees with the results of Ref. \cite{pacher03}.
%Fig. 2 {PRB10}

The super-radiant effect, as we have already mentioned in similar context above,
survives the intrinsic chaotic dynamics, in this case the diagonal site-disorder
\cite{VZWNMP04,PRB10}. For weak disorder, the maximum transmission occurs at
the same critical values of the continuum coupling. In the case of strong disorder,
with the width of site energy fluctuation exceeding the tunneling strength $v$,
the maximum transmission is shifted to higher values of the ratio $\gamma/v$.
The width average reveals two super-radiant transitions at the expected values of
the continuum coupling, $\gamma=2v$ and $\gamma=2qv$. The details of this dynamics
are discussed in ref. \cite{PRB10}, along with the comparison to the one-dimensional
Anderson model \cite{beenakker97,ida90,melsen95} with symmetric and asymmetric leads.

Coupled quantum dots have been already studied and used in the experiments directed to
the development of quantum informatics \cite{hanson07,kitavittaya09}. The effects similar
to what have been discussed in Sec. 4.1, including the decoupling of special states
from the leads that is equivalent to the appearance of bound states embedded in
the continuum were discovered in \cite{ladron03,ladron06}. There is a growing interest
to the geometry more complicated than a simple one-dimensional chain \cite{solomon08,
hsu09}. A three-dimensional transport diagram was probed \cite{granger10} in triple
quantum dots, where the conditions for stable transport were established. In a similar
direction the search was performed with $Y$-junctions of superconducting Josephson
chains \cite{giuliano09} and $T$-shaped quantum dot structures \cite{santos03}. Another
possible geometry corresponds to a tetrahedral qubit \cite{feigelman04}.

The theoretical results discussed above can be directly generalized to plane or
three-dimensional networks of various structures \cite{PRB10}. It was already noticed
\cite{giuliano09} that the $Y$-junctions behave similarly to quantum Brownian motion on
frustrated planar lattices. In the simplest models, continuum coupling is concentrated
at the perimeter. This arrangement can model an array of quantum dots
or a particle tunneling through a lattice, with or without local disorder.
At strong disorder, again the maximum transmission corresponds to the coupling
strength growing proportional to the degree of disorder. Especially interesting
can be the study of crossed chains that could be elements of future quantum computers
\cite{yan11}. The crossing point in such models plays
a singular role connecting all legs of the scheme. In addition to the Bloch
energy bands, here, even with no disorder, the states of different nature appear,
namely quasilocalized states with energy outside of the band and with wave functions
exponentially decaying from the central point (evanescent waves).

For the analysis of transport through realistic nanostructures it would be important
to study, theoretically and experimentally, the influence of external noise (or
``dephasing" \cite{cresti08}) in addition to internal disorder. One inevitable source of
noise is provided by electromagnetic fluctuations in the chain and in the background,
see for example \cite{kohler04,ischi05}. In our approach it can be modeled by adding
many weak incoherent channels. Another extremely important features which should be
introduced in the consideration are the effects of magnetic field.

\newpage

\section{Conclusion and outlook}

At the time of growing experimental and theoretical interest in open and marginally
stable mesoscopic systems, it is useful to take a broad view of the whole field
and find out the most general ideas and methods which would make possible to work
out a more or less unified physical picture of various specific examples. The approach
related to the effective non-Hermitian Hamiltonian is quite general and at the same time
easily adaptable to different cases.

This approach, being in its starting formulation a rigorous application of well known
quantum-mechanical ideas, keeps its generality and at the same time allows one to make
proper approximations for concrete problems. The main new feature stressed by this approach
is a collective interaction of internal states in a complex quantum system through
their common continuum or radiation field. Being first suggested in quantum optics,
this property turns out to be generic for all open quantum systems with many degrees of freedom.
The process of such interactions can be traced all the way from almost independent
well isolated resonances to super-radiance and trapping. Paradoxically, if the number
of open channels does not increase too fast, the typical widths of narrow resonances
become even narrower with excitation energy increasing. We tried to demonstrate the unified
physical picture of such phenomena in various subfields and such different objects
as nanostructures and exotic nuclei, complex atoms and hypothetical long-lived
heavy particles. The nuclear Ericson fluctuations turn out to be almost identical
to universal conductance fluctuations in quantum dots; there are suggestions to look
for similar physics even in such classical events as seismic waves.

The unified formulation of the super-radiance physics of open systems allows physicists
to understand better many questions which historically were asked and seemingly
solved on different grounds and for different occasions. The segregation of broad and narrow
resonances was first just an observation in complicated numerical simulations of nuclear
reactions which seemed to be infinitely remote from coherent spontaneous radiation in
the gas of two-level atoms. The details of Ericson fluctuations in the regime of
overlapping resonances were not confirmed by the experiments $-$ now we understand that
the phenomenon of super-radiance (or perfect coupling to the leads in mesoscopic physics)
related to the requirements of unitarity were not fully satisfied in the original formulation.
We anticipate many such partial discoveries in further applications.

As for any alive physical theory, we still have many open questions which hopefully
will be addressed in the future. Our shortest list would include

\begin{itemize}

\item clarification of the important problem of the role of internal chaos and interrelation
between the regular or chaotic dynamics inside the system and its openness properties;

\item role of the multitude of weak external channels present in many practical cases,
decoherence and equilibrium in open mesoscopic systems;

\item better understanding of interplay between collectivity in intrinsic interactions,
including superconducting correlations, and collectivity of super-radiant type through
the continuum;

\item quantum signal transmission in complicated geometry;

\item development of more powerful computational methods.

\end{itemize}

{\sl ACKNOWLEDGEMENTS}. Some of the original ideas were worked out together
with V.V. Sokolov. We are grateful to our coauthors on the latest stages of common
work, G.P. Berman, G.L. Celardo, F.M. Izrailev, L. Kaplan, R.A. Sen'kov, A.M. Smith,
and S. Sorathia for friendly and fruitful collaboration. A. Volya made
a great contribution to the discussed ideas and their practical implementation.
We acknowledge useful discussions with V.V. Flambaum, I. Rotter and H.A. Weidenm\"{u}ller.
The work was partially supported by the National Superconducting Cyclotron Laboratory
at Michigan State University, NSF through the grants PHY-0244453, PHY-0555366 and
PHY-0758099, and USA-Israel Binational Science Foundation.

\newpage

\begin{figure}
\includegraphics[width=120mm]{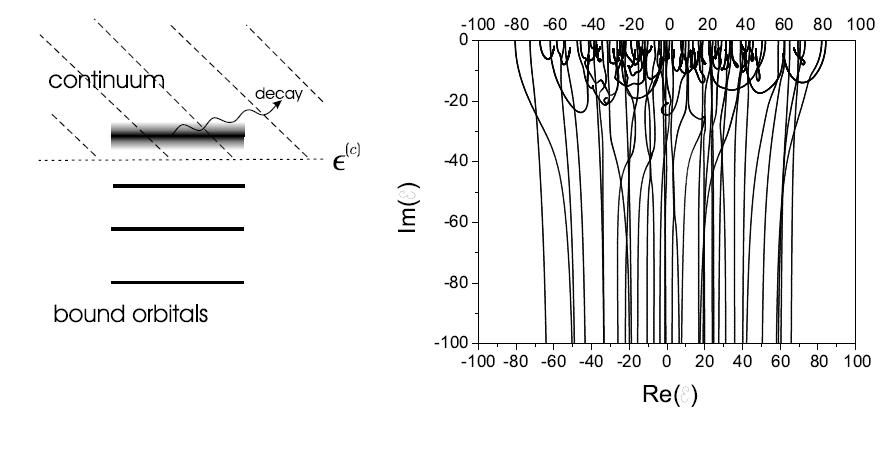}
\caption{A system of 4 fermions in 8 single-particle levels
[$8!/(4!4!)=70$ many-body states] with the upper orbital in the continuum.
Particles interact via random Hermitian interaction. The lines (right panel)
show the trajectories of many-body energies in the complex plane as the
width of the upper single-particle level increases; $7!/(4!3!)$ many-body
states become trapped and return to the vicinity of the real axis \cite{VZWNMP04}.}
\end{figure}

\begin{figure}
\includegraphics[width=120mm]{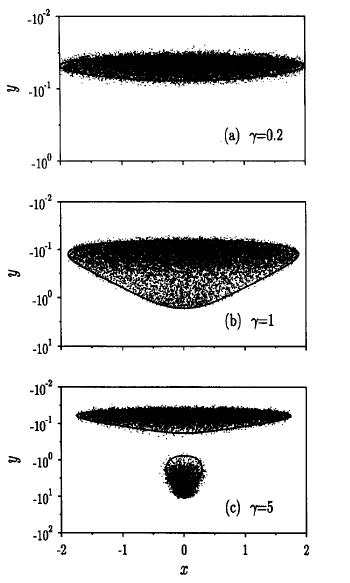}
\caption{ Segregation of the cloud of super-radiant states in the
complex energy plane $z=x+iy$ as coupling to continuum increases; here
$M/N=0.25$ \cite{lehmann95}. Note the logarithmic scale on the
vertical axis.}
\end{figure}

\begin{figure}
\includegraphics[width=120mm]{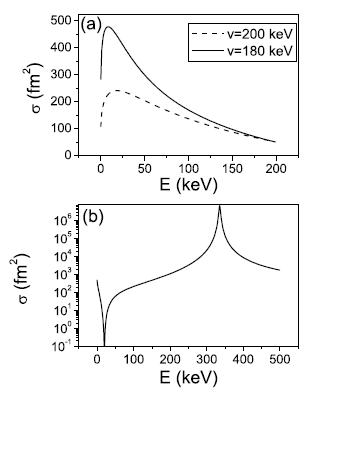}
\caption{The near-threshold scattering cross section in the two-level
model \cite{VZcont03}, see text. The parameters of eq. (\ref{64}) are selected
from comparison with a corresponding potential model. }
\end{figure}

\begin{figure}
\includegraphics[width=120mm]{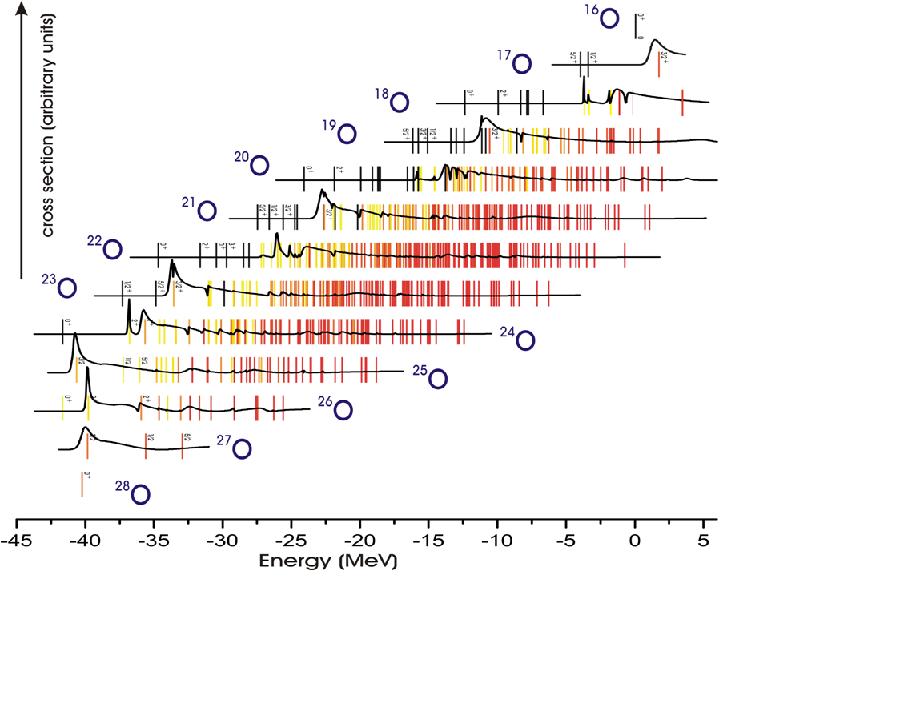}
\caption{Continuous shell model calculations for the chain of oxygen
isotopes \cite{VZCSM05}. Bound states, resonances and cross sections
of neutron scattering off the daughter nucleus are consistently calculated in
the unified approach. }
\end{figure}

\begin{figure}
\includegraphics[width=120mm]{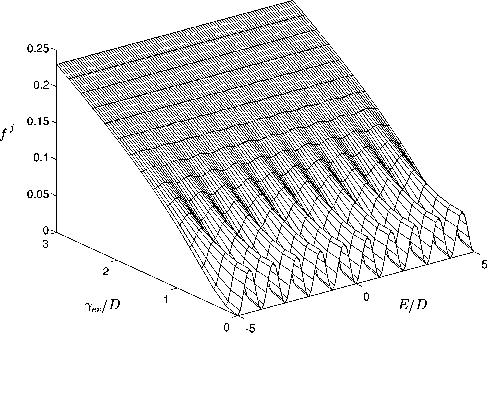}
\caption{The relative probability of excitation of the doorway state
from the continuum channel as a function of energy, $E/D$, and the
evaporation width, $\gamma_{{\rm ev}}/D$, of the background states \cite{SZsimple}
for a specific value of the spreading width, $\Gamma^{\downarrow}/D=10$.
The oscillations along the energy axis rapidly disappear as $\gamma_{{\rm ev}}$
grows.}
\end{figure}

\begin{figure}
\includegraphics[width=120mm]{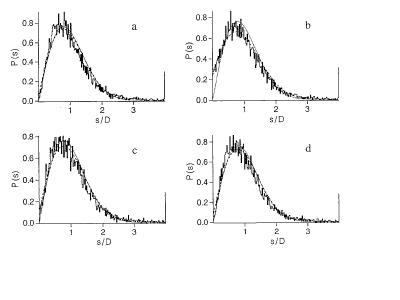}
\caption{ Nearest level spacing distribution found \cite{mizutori93}
by averaging 50 random matrices $160\times 160$ with the Gaussian distribution of
uncorrelated matrix elements and one open channel. The continuum coupling
parameter $\kappa=$0.1 (panel {\sl a}), 1 (panel {\sl b}), 5 (panel {\sl c}), and 50
(panel {\sl d}). At critical coupling, panel {\sl b}, the level repulsion at short
distances disappears and the distribution can be fitted as a combination
$\alpha P_{W}(s)+(1-\alpha)P_{G}(s)$ of the Wigner-Dyson distribution
$P_{W}(s)$ for the Gaussian orthogonal ensemble (GOE) and the Gaussian distribution
$P_{G}(s)$ centered at the mean value of $s$. After the separation of the
super-radiant state, the distribution returns to the GOE regime.}
\end{figure}

\begin{figure}
\includegraphics[width=120mm]{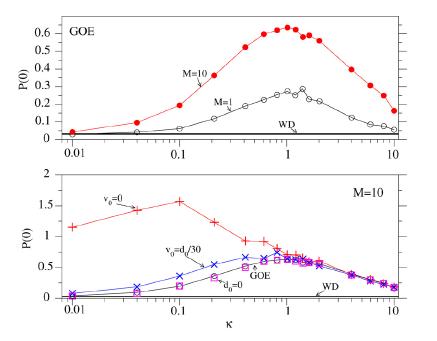}
\caption{ The probability $P(0)$ of finding the resonance energy spacing (in units
of the mean spacing) $s<0.04$ \cite{celardo08}. The intrinsic dynamics
corresponds to the GOE, upper panel, with different numbers of open channels $M$,
and to the TBRE, lower panel, for $M=10$, and different strength of the two-body
random interaction, from $v=0$ through the critical value for onset of intrinsic
chaos to the strong interaction for degenerate single-particle levels, when
the results are identical to those for the GOE case. Note that for the TBRE
the vertical scale is different; at weak interaction the deviations from $P(0)=0$
start at a very small continuum coupling strength.}
\end{figure}

\begin{figure}
\includegraphics[width=120mm]{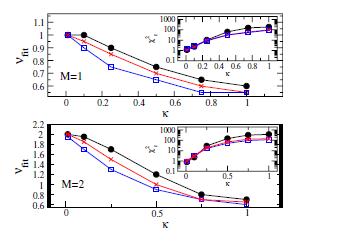}
\caption{The best fitted parameter $\nu$ for the chi-square distribution
with $\nu$ degrees of freedom used for the description of the width distribution
found with averaging over 1000 random realizations of intrinsic interactions for
one, upper panel, and two, lower panel, open channels \cite{celardo08}. Full
circles refer to the GOE intrinsic Hamiltonian, crosses stand for the TBRE with
the interaction below onset of chaos, and squares for TBRE with no intrinsic chaos.
The inserts show the unsatisfactory growing chi-square criterion of the fit as a function
of $\kappa$.}
\end{figure}

\begin{figure}
\includegraphics[width=120mm]{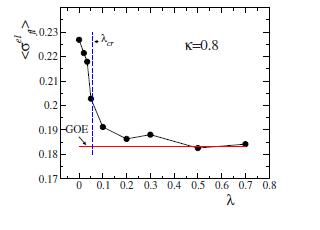}
\caption{ The fluctuating part of the elastic cross section for 10 open
channels and $\kappa=0.8$ as a function of the intrinsic interaction strength in
the TBRE; the GOE value is shown by a horizontal line \cite{celardo07}. The dashed
vertical line shows the critical value $\lambda_{{\rm cr}}$ for the transition to
chaos in the TBRE. }
\end{figure}

\begin{figure}
\includegraphics[width=120mm]{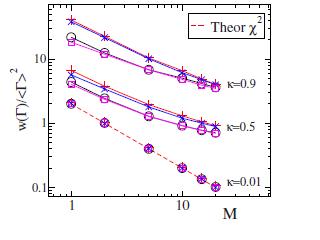}
\caption{ Normalized variance of the width as a function of the number of
channels, for different coupling strengths $\kappa$ \cite{celardo07}. Circles refer
to the GOE case, pluses to $\lambda=0$, crosses to $\lambda=1/30$, and squares to
$\lambda\rightarrow\infty$. While for weak coupling, $\kappa=0.01$, the variance
decreases with the number of channels very fast in accordance with the expected
chi-square distribution (dashed line), for strong couplings, $\kappa=0.5$ and 0.9,
the behavior is different from the $1/M$ dependence.}
\end{figure}

\begin{figure}
\includegraphics[width=120mm]{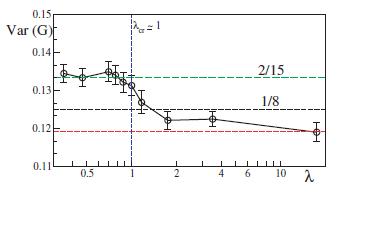}
\caption{Variance of the conductance, eq. (\ref{107}), for the TBRE model
(7 fermions in 14 orbitals with 20 open channels) as a function of the strength of
intrinsic interaction \cite{sorathia09}. }
\end{figure}

\begin{figure}
\includegraphics[width=120mm]{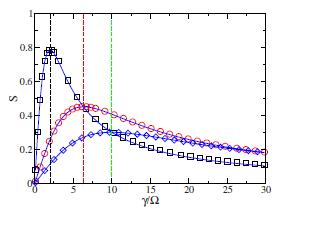}
\caption{ Integrated transmission, eq. (\ref{112}), for $N=100$ sites and
different asymmetry parameters $q$ \cite{PRB10}: squares refer to $q=1$,
circles to $q=10$, and diamonds to $q=25$. The vertical dashed lines indicate
the critical value of the coupling strength in each case, where the integrated
transmission is predicted to have its maximum.}
\end{figure}

\end{document}